\documentclass[11pt,a4paper]{article}

\usepackage[truedimen,margin=32mm]{geometry} 

\usepackage{mathrsfs}
\usepackage{amssymb}
\usepackage{amsmath}
\usepackage{ascmac}
\usepackage{amsthm}
\usepackage[pdftex]{graphicx}
\usepackage{natbib}
\usepackage{color}
\usepackage{setspace}
\usepackage{url}
\usepackage{here}
\usepackage{comment,booktabs}
\usepackage{lscape}
\usepackage{multirow}
\usepackage{algorithm, algorithmicx, algpseudocode}

\usepackage{titlesec}
\titleformat*{\section}{\large\bfseries}
\titleformat*{\subsection}{\it}

\theoremstyle{plain}

\theoremstyle{definition}
\newtheorem{rem}{Remark}

\newcommand{\diag}{\mathrm{diag}}
\newcommand{\E}{\mathrm{E}}

\title{{\bf 
Fast and Locally Adaptive Bayesian Quantile Smoothing 
using Calibrated Variational Approximations
}}
\date{}

\allowdisplaybreaks
\begin{document}

\maketitle
\doublespacing

\vspace{-1.5cm}
\begin{center}
{\large Takahiro Onizuka$^{1}$, Shintaro Hashimoto$^{1}$ and Shonosuke Sugasawa$^{2}$}
\end{center}

\medskip
\noindent
$^{1}$Department of Mathematics, Hiroshima University\\
$^{2}$Faculty of Economics, Keio University

\medskip
\medskip
\medskip
\begin{center}
{\bf \large Abstract}
\end{center}

Quantiles are useful characteristics of random variables that can provide substantial information on distributions compared with commonly used summary statistics such as means. In this study, we propose a Bayesian quantile trend filtering method to estimate the non-stationary trend of quantiles. We introduce general shrinkage priors to induce locally adaptive Bayesian inference on trends and mixture representation of the asymmetric Laplace likelihood. To quickly compute the posterior distribution, we develop calibrated mean-field variational approximations to guarantee that the frequentist coverage of credible intervals obtained from the approximated posterior is a specified nominal level. Simulation and empirical studies show that the proposed algorithm is computationally much more efficient than the Gibbs sampler and tends to provide stable inference results, especially for high/low quantiles.


\newpage
\section{Introduction}

Smoothing or trend estimation is an important statistical method to investigate characteristics of data, and such methods have been applied in various scientific fields such as 
astronomical spectroscopy \citep[e.g.][]{politsch2020trend}, 
biometrics \citep[e.g.][]{faulkner2020horseshoe}, 
bioinformatics \citep[e.g.][]{eilers2005quantile}, 
economics \citep[e.g.][]{yamada2022trend} and 
environmetrics \citep[e.g.][]{brantley2020baseline} among others. 
For estimating underlying trends, the $\ell_1$ trend filtering  \citep{kim2009ell_1,tibshirani2014adaptive} is known to be a powerful tool that can flexibly capture local abrupt changes in trends, compared with spline methods.
The $\ell_1$ trend filtering is known to be a special case of the generalized lasso proposed by \cite{tibshirani2011solution}. Furthermore, fast and efficient optimization algorithms for trend filtering have also been proposed \citep[e.g.][]{ramdas2016fast}. 
Due to such advantages in terms of flexibility and computation, extensions of the original trend filtering to spatial data \citep{wang2015trend} and functional data \citep{wakayama2021locally,wakayama2022functional} have been considered.
However, the majority of existing studies focus on estimating mean trends with a homogeneous variance structure, and these methods may not work well in the presence of outliers or data with heterogeneous variance. Additionally, we are often interested in estimating quantiles rather than means. 
Recently, \cite{brantley2020baseline} proposed a quantile version of trend filtering (QTF) by adding a $\ell_1$ penalization to the well-known check loss function, but the literature on quantile trend filtering remains scarce.

The main difficulty in applying the optimization-based trend filtering as considered in \cite{brantley2020baseline} is that uncertainty quantification for trend estimation is not straightforward.
Moreover, the frequentist formulation includes tuning parameters in the regularization, but the data-dependent selection of the tuning parameter is not obvious, especially with quantile smoothing. 
A reasonable alternative is to employ a Bayesian formulation for trend filtering by introducing priors. 
In particular, \cite{roualdes2015bayesian} and \cite{faulkner2018locally} proposed the use of shrinkage priors for differences between parameters and \cite{kowal2019dynamic} also considered a Bayesian formulation based on a dynamic shrinkage process under Gaussian likelihood to estimate the mean trend. 
However, the existing approach suffers mainly from three problems; 1) The current methods cannot be applied to quantile smoothing, 2) The posterior computation could be computationally intensive for a large sample size, and 3) The current methods cannot avoid the bias induced by model misspecification, therefore the resulting credible interval may not be valid.

In this study, we first proposed a Bayesian quantile trend filtering method. To this end, we employed the asymmetric Laplace distribution as a working likelihood \citep{yu2001bayesian, sriram2013posterior}. Combining the data augmentation strategy by \cite{kozumi2011gibbs}, we then constructed an efficient Gibbs sampling algorithm and a mean-field variational Bayes (MFVB) algorithm under two types of shrinkage priors; Laplace \citep{park2008bayesian} and horseshoe \citep{carvalho2010horseshoe} priors. 
It is well-known that the variational Bayes method enables the quick calculation of point estimates, while the MFVB algorithm tends to provide narrower credible intervals than that of Gibbs sampling \citep[e.g.][]{blei2017variational}. Additionally, the (possibly) misspecified asymmetric Laplace likelihood may produce invalid credible intervals. To overcome such problems, we proposed a new simulation-based calibration algorithm for variational posterior distribution which is expected to provide fast and valid credible intervals. 
We demonstrated the usefulness of the proposed methods through extensive simulation studies and real data examples.

The remainder of the paper is structured as follows: In Section \ref{sec:2}, we formulate a Bayesian quantile trend filtering method and provide Gibbs sampling and variational Bayes algorithms. In Section \ref{sec:3}, we describe the main proposal of this paper, a new calibration algorithm for approximating posterior distribution with variational Bayes approximation. In Section \ref{sec:4}, we illustrate simulation studies to compare the performance of the proposed methods. In Section \ref{sec:5}, we apply the proposed methods to real data examples. 
Concluding remarks are presented in Section~\ref{sec:conc}.
Additional information on the proposed algorithms and numerical experiments are provided in the Supplementary Material. R code implementing the proposed methods is available at GitHub repository (\url{https://github.com/Takahiro-Onizuka/BQTF-VB}).

\section{Bayesian quantile trend filtering}
\label{sec:2}

\subsection{Trend filtering via optimization}
\label{subsec:2.1}

Let $y_i = \theta_i +\varepsilon_i\quad (i=1,\dots,n)$ be a sequence model, where $y_i$ is an observation, $\theta_i$ is a true location and $\varepsilon_i$ is a noise. 
The estimate of $\ell_1$ trend filtering \citep{kim2009ell_1} is given by solving the optimization problem 
\begin{align}\label{freq_TL}
\hat{\theta}=\arg \min_{\theta}  \|y-\theta\|_2^2+\lambda\|D^{(k+1)}\theta\|_1 ,
\end{align}
where $y=(y_1,\dots,y_n)^{\top}$, $\theta=(\theta_1,\dots,\theta_n)^{\top}$, $D^{(k+1)}$ is a $(n-k-1)\times n$ difference operator matrix of order $k+1$, and $\lambda>0$ is a tuning constant.  
Depending on the different order $k$, we can express various smoothing such as piecewise constant, linear, quadratic, and so forth \citep[see e.g.][]{tibshirani2014adaptive}. 
A fast and efficient optimization algorithm for the problem \eqref{freq_TL} was proposed by \cite{ramdas2016fast}.

Recently, \cite{brantley2020baseline} proposed quantile trend filtering, defined as the optimization problem
\begin{align}\label{QTF}
\hat{\theta}_p=\arg \min_{\theta} \rho_p(y-\theta)+\lambda\|D^{(k+1)}\theta\|_1 ,
\end{align}
where $\lambda>0$ is a tuning constant and $\rho_p(\cdot)$ is a check loss function given by 
\begin{align}\label{check}
\rho_p(r)=\sum_{i=1}^nr_i\{ p-1(r_i<0) \}, \quad 0<p<1.
\end{align}
To solve the problem \eqref{QTF}, \cite{brantley2020baseline} proposed a parallelizable alternating direction method of multipliers (ADMM) algorithm, and also proposed the selection of smoothing parameters $\lambda$ using a modified criterion based on the extended Bayesian information criterion.

\subsection{Bayesian formulation and shrinkage priors for differences}
\label{subsec:2.2}

To conduct Bayesian inference of the quantile trend, we often use the following model:
\begin{align}
&y_i=\theta_{pi}+\varepsilon_i,\quad \varepsilon_i\sim \mathrm{AL}(p,\sigma^2), \quad i=1,\ldots,n, \label{ALseq}
\end{align}
where $\theta_{pi}$ and $\sigma^2$ are unknown parameters, $p$ is a fixed quantile level, and ${\rm AL}(p,\sigma^2)$ denotes the asymmetric Laplace distribution with the density function
\begin{align*}
f_{\mathrm{AL}(p)}(x)=\frac{p(1-p)}{\sigma^2}\exp\left\{-\rho_p\left(\frac{x}{\sigma^2}\right)\right\},
\end{align*}
where $p$ is a fixed constant which characterizes the quantile level, $\sigma^2$ (not $\sigma$) is a scale parameter, and $\rho_p(\cdot)$ is a check loss function defined by \eqref{check}. However, we often handle multiple observations per grid point in practice. Hereafter, we considered the following model which accounts for multiple observations per grid point:
\begin{align}\label{ALmodel}
&y_{ij}=\theta_p(x_i)+\varepsilon_{ij},\quad \varepsilon_{ij}\sim \mathrm{AL}(p,\sigma^2), \ \ \ i=1,\ldots,n, \quad j=1,\dots, N_i,
\end{align}
where $\theta_p(x)$ is a $p$-th quantile in the location $x$, $n$ is the number of locations data are observed, and $N_i$ is the amount of data for each location $x_i$. It is a natural generalization of the sequence model \eqref{ALseq} \citep[see also][]{heng2022bayesian}. 
Note that the model (\ref{ALmodel}) is a nonparametric quantile regression with a single covariate, and the proposed approach can easily be generalized for additive regression with multiple covariates. 
For simplicity in notation, we dropped the subscript $p$ from $\theta$ for the remainder of the paper.

We then introduce shrinkage priors on differences. 
We define the $(k+1)$th order difference operator $D$ as
\begin{align*}
    D=\left(
    \begin{matrix}
     I_{k+1} & O\\
     \multicolumn{2}{c}{D_n^{(k+1)}}
    \end{matrix}
    \right),
\end{align*}
where $I_{k+1}$ is $(k+1)\times (k+1)$ identity matrix, $O$ is zero matrix and $D_n^{(k+1)}$ is $(n-k-1)\times n$ standard difference matrix that is defined by
\begin{align*}
    D_n^{(1)}=\left(
    \begin{matrix}
     1 & -1 & &\\
     & \ddots& \ddots & \\
     & & 1&-1 
    \end{matrix}
    \right)\in \mathbb{R}^{(n-1)\times n},\quad 
    D_n^{(k+1)}=D_{n-k}^{(1)}D_n^{(k)}.
\end{align*}
We consider flexible shrinkage priors on $D\theta$, and the priors are  represented by 
\begin{align*}
D\theta\mid \tau^2,\sigma^2,w\sim N_n(0,\sigma^2 W) \ \  \text{with} \ \ W={\rm diag}(w_1^2,\ldots,w_{k+1}^2, \tau^2w_{k+2}^2,\dots,\tau^2w_n^2),
\end{align*}
where $w=(w_1,\ldots,w_n)$ represents local shrinkage parameters for each element in $D\theta$ and $\tau^2$ is a global shrinkage parameter. 
Since the $D$ is non-singular matrix, the prior of $\theta$ can be rewritten as
\begin{align}
    \theta\mid\tau^2,\sigma^2,w \sim N_n(0,\sigma^2(D^{\top}W^{-1} D)^{-1}).
    \label{prior--theta}
\end{align}
Note that since $(D\theta)_i=\theta_i\sim N(0,w_i^2)$ for $i=1,\dots,k+1$, $w_i$ ($i=1,\dots, k+1$) is not related to shrinkage of difference. For this reason, we assumed the conjugate inverse gamma distribution $\mathrm{IG}(a_{w_i},b_{w_i})$ for $w_i^2$. For $i=k+2,\dots,n$, we considered two types of distribution; $w_i\sim {\rm Exp}(1/2)$ and $w_i\sim C^{+}(0,1)$. These priors were motivated from Laplace or Bayesian lasso prior \citep{park2008bayesian} and horseshoe prior \citep{carvalho2010horseshoe}, respectively. 
Regarding the other parameters, we assigned $\sigma^2\sim {\rm IG}(a_{\sigma}, b_{\sigma})$ and $\tau \sim C^{+}(0, C_{\tau})$. The default choice of hyperparameters is $a_{\sigma}=b_{\sigma}=0.1$ and $C_{\tau}=1$. 

\begin{rem}\label{rem1}
We follow \cite{tibshirani2014adaptive} for discussion about an extension to the proposed method for the situation where data is observed at an irregular grid. It is equal so that the locations of data $x=(x_1,\dots,x_n)$ have the ordering $x_1<x_2<\dots<x_n$ and $d_j=x_{j+1}-x_j$ is not constant. 
This issue is related to nonparametric quantile regression. When the locations $x\in\mathbb{R}^n$ are irregular and strictly increasing, \cite{tibshirani2014adaptive} proposed an adjusted difference operator for $k\geq 1$
\begin{align*}
D_n^{(x,k+1)}=D_{n-k}^{(x,1)}\mathrm{diag}\left(\frac{k}{x_{k+1}-x_{1}},\dots,\frac{k}{x_n-x_{n-k}}\right)D_{n}^{(x,k)}
\end{align*}
where $D_{n}^{(x,1)}=D_n^{(1)}$ and when $x_1=1,x_2=2,\dots,x_n=n$, $D_n^{(x,k+1)}$ is equal to $D_n^{(k+1)}$ \citep[see also ][]{heng2022bayesian}. The matrix $D$ is given by
\begin{align*}
    D=\left(
    \begin{matrix}
     I_{k+1} & O\\
     \multicolumn{2}{c}{ D_n^{(x,k+1)}}
    \end{matrix}
    \right),
\end{align*}
where $O$ is a zero matrix.
\end{rem}

\subsection{Gibbs sampler}
\label{subsec:2.3}

We first derived a Gibbs sampler by using the stochastic representation of the asymmetric Laplace distribution \citep{kozumi2011gibbs}.
For $\varepsilon_{ij} \sim \mathrm{AL}(p,\sigma^2)$, we have the following stochastic expression: 
\[
\varepsilon_{ij}=\psi z_{ij}+\sqrt{\sigma^2 z_{ij} t^2}u_{ij},\quad \psi=\frac{1-2p}{p(1-p)},\quad t^2=\frac{2}{p(1-p)},
\]
where $u_{ij} \sim N(0,1)$ and $z_{ij}\mid \sigma^2 \sim \mathrm{Exp}(1/\sigma^2)$ for $i=1,\ldots,n$ and $j=1,\dots,N_i$. 
From the above expression, the conditional likelihood function of $y_{ij}$ is given by
\begin{align}
p(y_{ij}\mid \theta_i, z_{ij}, \sigma^2)=(2\pi t^2\sigma^2)^{-1/2}z_{ij}^{-1/2}\exp\left\{ -\frac{(y_{ij}-\theta_i-\psi z_{ij})^2}{2t^2\sigma^2z_{ij}}\right\}. \label{conditional likelihood}
\end{align}
Under the prior \eqref{prior--theta}, the full conditional distributions of $\theta$ and $z_i$ are given by 
\begin{align*}
&\theta\mid y,z,\sigma^2,\gamma^2
\sim N_n\left(A^{-1}B, \sigma^2 A^{-1}\right),\\
&z_{ij}\mid y_{ij},\theta_i,\sigma^2\sim \mathrm{GIG}\left(\frac{1}{2},\frac{( y_{ij}-\theta_i)^2}{t^2\sigma^2},\frac{\psi^2}{t^2\sigma^2}+\frac{2}{\sigma^2}\right), \ \ i=1,\ldots,n, \ j=1,\dots,N_i,
\end{align*}
respectively, where 
\begin{align*}
    A&=D^{\top}W^{-1}D+\frac{1}{t^2}\diag\left(\sum_{j=1}^{N_1}z_{1j}^{-1},\ldots,\sum_{j=1}^{N_n}z_{nj}^{-1}\right), \\
    B&=\left(\frac{1}{t^2}\sum_{j=1}^{N_1}\left(\frac{y_{1j}}{z_{1j}}-\psi\right) ,\dots,\frac{1}{t^2}\sum_{j=1}^{N_n}\left(\frac{ y_{nj}}{z_{nj}}-\psi\right)\right)^{\top}
\end{align*}
and $\mathrm{GIG}(a,b,c)$ denotes the generalized inverse Gaussian distribution.
The full conditional distribution of the scale parameter $\sigma^2$ is given by
\begin{align*}
\sigma^2\mid y,\theta, z, w,\tau^2\sim \mathrm{IG}\left(\frac{n+3N}{2}+a_{\sigma},
\alpha_{\sigma^2}
\right),
\end{align*}
where $N=\sum_{i=1}^n N_i$ is the number of observed values and
\begin{align*}
\alpha_{\sigma^2}=\sum_{i=1}^n\sum_{j=1}^{N_i}\frac{(y_{ij}-\theta_i-\psi z_{ij})^2}{2t^2z_{ij}}+\frac{1}{2}\theta^\top D^\top W^{-1}D\theta+\sum_{i=1}^n\sum_{j=1}^{N_i} z_{ij}+b_{\sigma}.
\end{align*}
By using the augmentation of the half-Cauchy distribution \citep{makalic2015simple}, the full conditional distribution of the global shrinkage parameter $\tau^2$ is given by
\begin{align*}
\tau^2\mid \theta, w, \sigma^2,\xi\sim \mathrm{IG}\left(\frac{n-k}{2}, \frac1{2\sigma^2}\sum_{i=k+1}^n\frac{\eta_i^2}{w_i^2}+\frac{1}{\xi}\right), \quad 
\xi\mid \tau^2\sim \mathrm{IG}\left(\frac{1}{2},\frac{1}{\tau^2}+1\right),
\end{align*}
where $\xi$ is an augmented parameter for $\tau^2$. For $i=1,\dots,k+1$, we assumed the prior $\mathrm{IG}(a_{w_i}, b_{w_i})$ for $w_i$, and then the full conditional distribution of $w_i$ is given by
\begin{align*}
    w_i^2\mid \theta, \sigma^2
    \sim \mathrm{IG}\left(\frac{1}{2}+a_{w_i}, \frac{\eta_i^2}{2\sigma^2}+b_{w_i}\right), \ \ i=1,\dots, k+1.
\end{align*}
The full conditional distribution of local shrinkage parameter $w_i$ ($i=k+2,\dots,n$) depends on the choice of the prior, either Laplace or horseshoe prior. 

\begin{itemize}
\item[-]{\bf (Laplace-type prior)}  \ \ 
The full conditional distributions of $\theta$, $z_i$, and $\sigma^2$ were already derived. For Laplace-type prior, we set $\tau^2=1$ and assume $w_i\mid \gamma^2 \sim \mathrm{Exp}(\gamma^2/2)$ for $i=k+2,\dots,n$. Then we have $(D\theta)_i \sim \mathrm{Lap}(\gamma)$. Noting that $\gamma \sim C^+(0,1)$, sampling from the standard half-Cauchy prior is equivalent to $\gamma^2 \mid \nu \sim \mathrm{IG}(1/2,1/\nu)$ and $\nu\sim \mathrm{IG}(1/2, 1)$ using the augmentation technique \citep{makalic2015simple}. Hence, the full conditional distributions of $w_i$, $\gamma^2$ and $\nu$ are, respectively, given by 
\begin{align*}
&w_i^2\mid \theta,\sigma^2,\gamma^2 \sim \mathrm{GIG}\left(\frac{1}{2},\frac{\eta_i^2}{\sigma^2},\gamma^2\right),\\
&\gamma^2\mid w, \nu\sim \mathrm{GIG}\left(n-k-\frac{3}{2},\frac{2}{\nu},\sum_{i=k+2}^n w_i^2\right),\quad \nu\mid \gamma^2 \sim \mathrm{IG}\left(\frac{1}{2},\frac{1}{\gamma^2}+1\right).
\end{align*}

\item[-]{\bf (Horseshoe-type prior)} \ \ 
The full conditional distributions of $\theta$, $z_i$, $\sigma^2$ and $\tau^2$ were already derived. For Horseshoe-type prior, we assume $w_i \sim C^+(0,1)$ for $i=k+2,\dots,n$. By using the representation $w_i^2\mid \nu_i \sim \mathrm{IG}(1/2,1/\nu_i)$ and $\nu_i\sim \mathrm{IG}(1/2, 1)$, the full conditional distributions of $w_i$ and $\nu_i$ are, respectively, given by 
\begin{align*}
&w_i^2\mid \theta,\sigma^2,\tau^2,\nu_i \sim \mathrm{IG}\left(1,\frac{1}{\nu_i}+\frac{\eta_i^2}{2\sigma^2\tau^2}\right),\quad \nu_i\mid w_i \sim \mathrm{IG}\left(\frac{1}{2},\frac{1}{w_i^2}+1\right).
\end{align*}

\end{itemize}

\subsection{Variational Bayes approximation}\label{subsec:2.4}

The MCMC algorithm presented in Section~\ref{subsec:2.3} can be computationally intensive when the sample size is large. 
For the quick computation of the posterior distribution, we derived the variational Bayes approximation \citep[e.g.][]{blei2017variational,tran2021practical} of the joint posterior. 
The idea of the variational Bayes method is to approximate an intractable posterior distribution by using a simpler probability distribution. Note that the variational Bayes method does not require sampling from the posterior distribution like MCMC, and it searches for an optimal variational posterior by using the optimization method. 
In particular, we employed the mean-field variational Bayes (MFVB) approximation algorithms that require the forms of full conditional distributions as given in Section~\ref{subsec:2.3}.

The variational distribution $q^*(\theta)\in \mathcal{Q}$ is defined by the minimizer of the Kullback-Leibler divergence from $q(\theta)$ to the true posterior distribution $p(\theta \vert y)$
\begin{align}
    q^*=\arg \min_{q\in \mathcal{Q}}
   \mathrm{KL}(q \| p(\cdot\mid y))
   =\arg \min_{q\in \mathcal{Q}}
   \int q(\theta)\log \frac{q(\theta)}{p(\theta\mid y)}d\theta. \label{KL}
\end{align}
If $\theta$ is decomposed as $\theta=(\theta_1,\ldots,\theta_K)$ and parameters $\theta_1,\ldots,\theta_K$ are mutually independent, each variational posterior can be updated as 
\begin{align*}
q(\theta_k)\propto  
\exp(\mathrm{E}_{\theta_{-k}}[\log p(y,\theta)])=\exp\left(\int q(\theta_k)\log p(y,\theta)d\theta_{-k}\right), \ \ \ \ k=1,\ldots,K,
\end{align*}
where $\theta_{-k}$ denotes the parameters other than $\theta_k$ and $\mathrm{E}_{\theta_{-k}}[\cdot]$ denotes the expectation under the probability density given parameters except for $\theta_k$.
Such a form of approximation is known as the MFVB approximation. 
If the full conditional distribution of $\theta_k$ has a familiar form, the above formula is easy to compute. 
According to the Gibbs sampling algorithm in Section \ref{subsec:2.3}, we used the following form of variational posteriors: 
\begin{align*}
q(\theta, z, \sigma^2, \tau^2, \xi)=q(\theta)q(z)q(\sigma^2)q(\tau^2)q(\xi),
\end{align*}
where 
\begin{align}
\label{vb-theta-z-sigma-tau-xi}
\begin{split}
&q(\theta) \sim N_n(A^{-1}B,(E_{1/\sigma^2}A)^{-1}), \ \ \ \ q(z_{ij})\sim \mathrm{GIG}\left(\frac{1}{2},\alpha_{z_{ij}},\beta_{z_{ij}}\right),\\
&q(\sigma^2)\sim \mathrm{IG}\left(\frac{n+3N}{2}+a_{\sigma},\alpha_{\sigma^2}\right), \ \ \ \ 
q(\tau^2) \sim \mathrm{IG}\left(\frac{n-k}{2},\alpha_{\tau^2}\right),\\
&q(\xi)\sim \mathrm{IG}\left(\frac{1}{2},E_{1/\tau^2}+1\right).
\end{split}
\end{align}
For $i=1,\dots,k+1$, we assume the prior $\mathrm{IG}(a_{w_i}, b_{w_i})$ for $w_i$, and then the variational distribution of $w_i$ is  given by
\begin{align*}
    q(w_i^2)
    \sim \mathrm{IG}\left(\frac{1}{2}+a_{w_i}, \frac{1}{2}E_{\eta_i^2}E_{1/\sigma^2}+b_{w_i}\right).
\end{align*}
The variational distributions of the other parameters depended on the specific choice of the distributional form of $\pi(w_i)$ $(i=k+2,\dots,n)$, which are provided as follows: 

\begin{itemize}
\item[-]{\bf (Laplace-type prior)}  \ \ 
The variational distributions for $w_i^2$ ($i=k+2,\dots,n$), $\gamma^2$ and $\nu$ are given by
\begin{align*}
&q(w_i^2) \sim \mathrm{GIG}\left(\frac{1}{2}, \alpha_{w_i^2}, E_{\gamma^2}\right),\\
&q(\gamma^2) \sim \mathrm{GIG}\left(n-k-\frac{3}{2}, 2E_{1/\nu}, \sum_{i=k+2}^nE_{w_i^2}\right),
\quad q(\nu) \sim \mathrm{IG}\left(\frac{1}{2},E_{1/\gamma^2}+1\right),
\end{align*}

\item[-]{\bf (Horseshoe-type prior)} \ \ 
The variational distributions for $w_i^2$ and $\nu_i$ ($i=k+2,\dots, n$) are given by
\begin{align*}
&q(w_i^2)\sim \mathrm{IG}(1,\alpha_{w_i^2}),\quad q(\nu_i)\sim \mathrm{IG}\left(\frac{1}{2},E_{1/w_i^2}+1\right),
\end{align*}

\end{itemize}

To obtain the variational parameters in each distribution, we update the parameters by using the coordinate ascent algorithm \citep[e.g.][]{blei2017variational}. The two proposed variational algorithms based on the above variational distributions are given in Algorithm \ref{algo:VB-Lap} and \ref{algo:VB-HS}. Note that we set $\epsilon=10^{-4}$ as the convergence criterion in the simulation study, $e_i$ is a unit vector that the $i$th component is 1,  $d_i^{\top}$ is the $i$th row of difference matrix $D$, and  $K_c(\cdot)$ is the modified Bessel function of the second kind with order $c$ in Algorithms \ref{algo:VB-Lap} and \ref{algo:VB-HS}.

\begin{algorithm*}[thbp]
\caption{\bf--- \small Variational Bayes approximation under Laplace prior.}
\label{algo:VB-Lap}
{
Initialize: $E_{z_{ij}}, E_{1/z_{ij}}, E_{1/w_i}, E_{1/\sigma^2}, E_{\gamma^2}, E_{1/\nu}>0$ $(j=1,\dots,N_i, i=1,\dots,n)$. Set $E_{1/\tau^2}=1, E_{1/\xi}=0$ under Laplace prior. 
\begin{itemize}
\item[1.] Cycle the following:
{\footnotesize
\begin{align*}
\mathrm{(i)}\quad &A\leftarrow\frac{1}{t^2}
\diag\left(\sum_{j=1}^{N_1}E_{1/z_{1j}},\dots,\sum_{j=1}^{N_n}E_{1/z_{nj}}\right)+D^{\top}\hat{W}^{-1}D,\\
&B\leftarrow\frac{1}{t^2}\left(C-\psi 1_n\right),\quad C\leftarrow\left(\sum_{j=1}^{N_1}y_{1j}E_{1/z_{1j}},\dots, \sum_{j=1}^{N_n}y_{nj}E_{1/z_{nj}}\right)^{\top},\\
&\hat{W}^{-1}\leftarrow\diag(E_{1/w_1^2},\dots,E_{1/w_{k+1}^2},E_{1/\tau^2}E_{1/w_{k+2}^2},\dots,E_{1/\tau^2}E_{1/w_n^2}),\\ &E_{\theta_i}\leftarrow(A^{-1}B)_i, \quad E_{\theta_i^2}\leftarrow e_i^{\top}(E_{\sigma^2}^{-1}A^{-1}+A^{-1}BB^{\top}A^{-1})e_i,\\
&E_{\eta_i^2}\leftarrow  d_i^{\top}(E_{\sigma^2}^{-1}A^{-1}+A^{-1}BB^{\top}A^{-1})d_i \quad  (i=1,\dots,n),\\
&\alpha_{\sigma^2}\leftarrow\frac{1}{2t^2}\sum_{i=1}^n\sum_{j=1}^{N_i}\left(y_{ij}^2E_{1/z_{ij}}-2\psi y_{ij}+\psi^2E_{z_{ij}}-2(E_{1/z_{ij}}y_{ij}-\psi)E_{\theta_i}+E_{\theta_i^2}E_{1/z_{ij}}\right)\\
& +\frac{1}{2}\sum_{i=1}^{k+1}E_{\eta_i^2}E_{1/w_i^2}+\frac{1}{2}\sum_{i=k+2}^{n}E_{\eta_i^2}E_{1/w_i^2}E_{1/\tau^2}+\sum_{i=1}^n\sum_{j=1}^{N_i}E_{z_{ij}}+b_{\sigma},\\
&E_{\sigma^2}\leftarrow\frac{2\alpha_{\sigma^2}}{n+3N+2a_{\sigma}-2}, \quad E_{1/\sigma^2}\leftarrow\frac{n+3N+2a_{\sigma}}{2\alpha_{\sigma^2}},\\
&\alpha_{z_{ij}}\leftarrow\frac{1}{t^2}(y_{ij}^2-2y_{ij}E_{\theta_i}+E_{\theta_i^2})E_{1/\sigma^2},\quad \beta_{z_{ij}}\leftarrow\left(\frac{\psi^2}{t^2}+2\right)E_{1/\sigma^2},\\
&E_{z_{ij}}\leftarrow\frac{\sqrt{a_{z_{ij}}}K_{3/2}(\sqrt{a_{z_{ij}}b_{z_{ij}}})}{\sqrt{b_{z_{ij}}}K_{1/2}(\sqrt{a_{z_{ij}}b_{z_{ij}}})},\\
&E_{1/z_{ij}}\leftarrow\frac{\sqrt{b_{z_{ij}}}K_{3/2}(\sqrt{a_{z_{ij}}b_{z_{ij}}})}{\sqrt{a_{z_{ij}}}K_{1/2}(\sqrt{a_{z_{ij}}b_{z_{ij}}})}-\frac{1}{a_{z_{ij}}}\quad  (j=1,\dots,N_i, i=1,\dots,n),\\
& E_{1/w_i^2}\leftarrow(1+2a_{w_i})/(E_{\eta_i^2}E_{1/\sigma^2}+2b_{w_i})\quad (i=1,\dots,k+1)\\
\mathrm{(ii)}\quad &\alpha_{w_i^2}\leftarrow E_{1/\sigma^2}E_{\eta_i^2},\quad E_{w_i^2}\leftarrow
\frac{\sqrt{\alpha_{w_i^2}}K_{3/2}(\sqrt{\alpha_{w_i^2}E_{\gamma^2}})}{\sqrt{E_{\gamma^2}}K_{1/2}(\sqrt{\alpha_{w_i^2}E_{\gamma^2}})},\\
&E_{1/w_i^2}\leftarrow\frac{\sqrt{E_{\gamma^2}}K_{3/2}(\sqrt{a_{w_i^2}E_{\gamma^2}})}{\sqrt{a_{w_i^2}}K_{1/2}(\sqrt{a_{w_i^2}E_{\gamma^2}})}-\frac{1}{\alpha_{w_i^2}}\quad  (i=k+2,\dots,n),\\ &E_{\gamma^2}\leftarrow\frac{\sqrt{2E_{1/\nu}}K_{n-k-1/2}(\sqrt{2E_{1/\nu}\sum_{i=k+2}^nE_{w_i^2}})}{\sqrt{\sum_{i=k+2}^nE_{w_i^2}}K_{n-k-3/2}(\sqrt{2E_{1/\nu}\sum_{i=k+2}^nE_{w_i^2}})},\\
&E_{1/\gamma^2}\leftarrow\frac{\sqrt{\sum_{i=k+2}^nE_{w_i^2}}K_{n-k-1/2}(\sqrt{2E_{1/\nu}\sum_{i=k+2}^nE_{w_i^2}})}{\sqrt{2E_{1/\nu}}K_{n-k-3/2}(\sqrt{2E_{1/\nu}\sum_{i=k+2}^nE_{w_i^2}})},\\ &E_{1/\nu}\leftarrow\frac{1}{2(E_{1/\gamma^2}+1)}.
\end{align*}
}
\item[2.] For iteration $\ell$ in step 1 and convergence criterion $\epsilon>0$, if $\vert E_{\theta_i}^{(\ell)}-E_{\theta_i}^{(\ell-1)}\vert <\epsilon$, stop the algorithm.
\end{itemize}
}
\end{algorithm*}

\begin{algorithm*}[thbp]
\caption{\bf--- \small Variational Bayes approximation under horseshoe prior.}
\label{algo:VB-HS}

Initialize: $E_{z_{ij}}, E_{1/z_{ij}}, E_{1/w_i}, E_{1/\tau^2}, E_{1/\sigma^2}, E_{1/\xi^2}, E_{1/\nu_i}>0$ $(j=1,\dots,N_i, i=1,\dots,n)$.

\begin{itemize}
\item[1.] Cycle the following:
\begin{align*}
\mathrm{(i)}\quad &\mathrm{Same\ update\ as\ (i)\ in\ Algorithm\ 1.}\\
\mathrm{(ii)}\quad & \alpha_{w_i^2}\leftarrow E_{1/\nu_i}+\frac{1}{2}E_{1/\sigma^2}E_{1/\tau^2}E_{\eta_i^2},\\
&E_{1/w_i^2}\leftarrow\frac{1}{a_{w_i^2}},\quad 
E_{1/\nu_i^2}\leftarrow\frac{1}{2(E_{1/w_i^2}+1)}\quad (i=k+2,\dots,n),\\
&\alpha_{\tau^2}\leftarrow\frac{1}{2}\sum_{i=k+2}^n E_{\eta_i^2}E_{1/w_i^2}E_{1/\sigma^2}+E_{1/\xi},\quad E_{1/\tau^2}\leftarrow\frac{n-k}{2\alpha_{\tau^2}},\\
&E_{1/\xi}\leftarrow\frac{1}{2(E_{1/\tau^2}+1)}.
\end{align*}
\item[2.] For some iteration $\ell$ in step 1 and convergence criterion $\epsilon>0$, if $\vert E_{\theta_i}^{(\ell)}-E_{\theta_i}^{(\ell-1)}\vert <\epsilon$, stop the algorithm.
\end{itemize}

\end{algorithm*}

\section{Calibrated variational Bayes approximation}\label{sec:3}

The main proposal of this study is described below. 
When we use the mean field variational Bayes method, the posterior credible intervals are calculated based on the quantile of the variational posterior. In the proposed model, the variational distribution of the parameter of interest $\theta_i$ is represented by the normal distribution $N(\mu_i, \Sigma_{ii})$, where the mean $\mu_i$ and variance $\Sigma_{ii}$ are defined in Section~\ref{subsec:2.4}. Although the variational approximation provides the point estimate quickly, the corresponding credible interval tends to be narrow in general \citep[e.g.][]{wand2011mean, blei2017variational}. Additionally, it is well-known that the credible interval can be affected by model misspecification, as addressed by \cite{sriram2013posterior} and \cite{sriram2015sandwich} in the Bayesian linear quantile regression. 
Hence, if the asymmetric Laplace working likelihood in the proposed model is mis-specified, the proposed model would not have been able to provide valid credible intervals even if we use the MCMC algorithm.

As presented in the previous subsection, the conditional prior and likelihood of $\theta$ were given by \eqref{prior--theta} and \eqref{conditional likelihood}, respectively. Here we add a common (non-random) scale parameter $\lambda$, and then replace \eqref{prior--theta} and \eqref{conditional likelihood} with 
\begin{align*}
p(y_{ij}\mid \theta_i, z_{ij}, \sigma^2)&=(2\pi t^2\lambda\sigma^2)^{-1/2}z_{ij}^{-1/2}\exp\left\{ -\frac{(y_{ij}-\theta_i-\psi z_{ij})^2}{2t^2\lambda\sigma^2z_{ij}}\right\},\\
p(\theta\mid \sigma^2, \tau, w)&=(2\pi\lambda\sigma^2)^{-n/2} \vert D^{\top}W^{-1}D\vert^{1/2} \exp\left(-\frac{1}{2\lambda\sigma^2}\theta^{\top} D^{\top}W^{-1}D\theta \right),
\end{align*}
respectively. Based on these representations, the variational posterior of $\theta$ is given by
\begin{align*}
    q(\theta) \sim N_n(\mu, \lambda\Sigma).
\end{align*}
The constant $\lambda$ in the likelihood and conditional prior controls  the scale of the variational posterior. Indeed, it is natural that the scale of the posteriors was determined by the scale of the likelihood and prior. If the scale parameter $\lambda$ is given locally for each $\theta_i$ (i.e. $\lambda_i$), then the variational posterior of $\theta_i$ is also given by $q(\theta_i) \sim N_n(\mu_i, \lambda_i\Sigma_{ii})$ for each $i$. 
We used the formulation to calibrate credible intervals after the point estimation. The proposed calibration algorithm is given in Algorithm~\ref{algo-lambda}.

\begin{algorithm*}[t]
\caption{\bf--- Calibration of variational posterior.}

For calibration of variational posterior at the quantile level $p$, we set the monotonically increasing sequence $1=\lambda_1<\lambda_2<\dots$ and run the following four steps:
\begin{enumerate}

\item Estimate the variational posterior for $p$-th quantile trend filtering $q(\theta)\approx N_n(\mu^*, \Sigma^*)$ using the observed data $y=(y_1,\dots,y_n)$. 

\item Run the variational algorithm and estimate the variational posterior for $0.5$-th quantile $q(\theta)\approx N_n(\mu_{50\%}, \Sigma_{50\%})$ using the observed data $y$. 

\item Generate $B$ bootstrap samples $y^{(1)},\dots,y^{(B)}$ based on the residuals $y-\mu_{50\%}$, and calculate the variational posteriors as $N_n(\mu^{(1)}, \Sigma^{(1)}), \dots, N_n(\mu^{(B)}, \Sigma^{(B)})$ using bootstrap samples $y^{(1)},\dots,y^{(B)}$.

\item Regarding $\{\mu^{(1)},\ldots,\mu^{(B)}\}$ as $B$ posterior samples, for $i=1,\dots,n$, evaluate the empirical coverage probability 
\begin{align*}
    \hat{c}_{\alpha, i}(\lambda_{\ell})=\frac{1}{B}\sum_{b=1}^B 1\{\mu_i^{(b)} \in C_{\alpha}(\mu_i^*, \lambda_{\ell}\Sigma_{ii}^*)\},
\end{align*}
where $C_{\alpha,i}(\mu_{i}^*, \lambda_{\ell}\Sigma_{ii}^*)$ is $100(1-\alpha)$\% credible intervals under $N(\mu_{i}^*, \lambda_{\ell}\Sigma_{ii}^*)$ $(\ell=1,2,\dots)$. Then, selecte the optimal value $\hat{\lambda}$ so that 
\begin{align*}
    \hat{\lambda}_i=\mathrm{argmin}_{\lambda} \{\hat{c}_{\alpha,i}(\lambda)-(1-\alpha)\},
\end{align*}
and $q(\theta_i) \sim N_n(\mu_i^*, \hat{\lambda}_{i}\Sigma_{ii}^*)$ is  the calibrated variational posterior distribution.
\end{enumerate}
\label{algo-lambda}
\end{algorithm*}

Algorithm~\ref{algo-lambda} is similar to the calibration method for general Bayes credible regions proposed by \cite{syring2019calibrating}, but the proposed algorithm drastically differs from the existing calibration method in that it computes variational Bayes posteriors for $B$ times while the calibration method by \cite{syring2019calibrating} runs MCMC algorithms for $B$ times. 
Thus, the proposed algorithm is computationally much faster than the existing calibration algorithm. 
Further, compared with the Gibbs sampler presented in Section~\ref{subsec:2.3}, steps 3 and 4 in Algorithm~\ref{algo-lambda} can be parallelized so that a significant reduction of computational costs can be attained with the proposed method.

After we obtain the optimal value of $\lambda$ using Algorithm \ref{algo-lambda}, we use $q(\theta) \sim N_n(\mu^*, \lambda\Sigma^*)$ as the calibrated variational posterior distribution. We then construct the calibrated credible interval of $\theta_i$ ($i=1,\dots,n$) by calculating the quantile of the marginal distribution of $N_n(\mu^*, \lambda\Sigma^*)$.
Here we used the residual bootstrap method \citep[e.g.][]{efron1982jackknife} to obtain bootstrap samples in Algorithm \ref{algo-lambda}. Since this algorithm is based on such semiparametric bootstrap sampling, robust uncertainty quantification can be expected even if the asymmetric Laplace assumption is violated.

\begin{rem}
In Algorithm \ref{algo-lambda}, we employed the residual bootstrap using 50\% quantile trend estimate as a fitted value when we estimated any quantile level. At first glance, it might seem like it is better to use bootstrap sampling based on residue $y-\mu_{p}$, where $\mu_p$ is $p$-th quantile trend estimate. 
However, since our aim is to re-sample from the empirical distribution of the original dataset $y$, the use of a 50\% quantile trend estimated as a fitted value in residual bootstrap is reasonable in practice. This is the critical point of Algorithm 1.
\end{rem}

To show the theoretical results of Algorithm \ref{algo-lambda} is not easy as well as the algorithm by \cite{syring2019calibrating} because it needs to evaluate the approximation errors of both bootstrap and variational approximations. We confirm the proposed algorithm through numerical experiments in the next section.

\section{Simulation studies}
\label{sec:4}

We illustrate the performance of the proposed method through simulation studies.

\subsection{Simulation setting}
\label{subsec:4.1}

To compare the performance of the proposed methods, we considered the following data generating processes \citep[see also][]{faulkner2018locally, brantley2020baseline}: We assumed that the data generating process was
\[y_i= f(x_i)+\varepsilon(x_i), \quad i=1,\dots,100,\]
where $f(x)$ is a true function and $\varepsilon(x)$ is a noise function. First, we considered the following two true functions: 

\begin{itemize}
\item[-] 
{\bf Piecewise constant (PC)}
\begin{align*}
f(x)&=2.5\cdot I(1\leq x\leq 20)+I(21\leq x\leq 40)\\
& \ \ \ \ 
+3.5\cdot I(41\leq x\leq 60)+1.5\cdot I(61\leq x\leq 100)
\end{align*}

\item[-] 
{\bf Varying smoothness (VS)}
\begin{align*}
f(x)=2+\sin(4x-2)+2\exp(-30(4x-2)^2).
\end{align*}
\end{itemize}
Since the scenario (PC) has three change points at $x=21$, $41$, and $61$, we aim to assess the ability to catch a constant trend and jumping structure. 
The second scenario (VS) is smooth and has a rapid change near $x=50$. Hence, the scenario is reasonable to confirm the shrinkage effect of the proposed methods and the adaptation of localized change. As noise functions, we considered the following three scenarios that represented the heterogeneous variance and various types of model misspecification.
\begin{itemize}
    \item[(I)] Gaussian noise: $\varepsilon(x)\sim N(0,\{(1+x^2)/4\}^2)$.
    \item[(II)] Beta noise: $\varepsilon(x)\sim \mathrm{Beta}(1,11-10x)$.
    \item[(III)] Mixed normal noise: $\varepsilon(x)\sim x N(-0.2,0.5)+(1-x) N(0.2,0.5)$.
\end{itemize}
For each scenario, simulated true quantile trends are summarized in Figure S1 of the Supplementary Materials. True quantile trends were computed from the quantiles of point-wise noise distributions. 
We next introduce the details of simulations. We used the two MCMC methods (denoted by MCMC-HS and MCMC-Lap), two non-calibrated variational Bayes methods (denoted by VB-HS and VB-Lap), and two calibrated variational Bayes methods (denoted by CVB-HS and CVB-Lap), where HS and Lap are the horseshoe and Laplace priors, respectively.
Note that we implemented CVB without parallelization although the bootstrap calibration steps in CVB can be parallelized.
To compare with the frequentist method, we used the quantile trend filtering based on the ADMM algorithm proposed by \cite{brantley2020baseline}, where the penalty parameter of Brantley's method was determined by the extended Bayesian information criteria. The method can be implemented using their {\tt R} package in  \url{https://github.com/halleybrantley/detrendr}.
For the order of trend filtering, we considered $k=0$ for (PC) and $k=1$ for (VS). Note that $k=0,1$ express the piecewise constant and the piecewise linear, respectively. We generated 7,500 posterior samples by using the Gibbs sampler presented in Section \ref{subsec:2.3}, and then only every 10th scan was saved (thinning). As criteria to measure the performance, we adopted the mean squared error (MSE), mean absolute deviation (MAD), mean credible interval width (MCIW), and coverage probability (CP) which are defined by
\begin{align*}
    &\mathrm{MSE}=\frac{1}{n}\sum_{i=1}^n (\theta_i-\hat{\theta}_i)^2,\quad \mathrm{MAD}=\frac{1}{n}\sum_{i=1}^n \vert \theta_i-\hat{\theta}_i\vert ,\\ &\mathrm{MCIW}=\frac{1}{n}\sum_{i=1}^n \hat{\theta}_{97.5,i}-\hat{\theta}_{2.5,i},\quad
    \mathrm{CP}=\frac{1}{n}\sum_{i=1}^n I(\hat{\theta}_{2.5,i}\leq \theta_i^* \leq \hat{\theta}_{97.5,i}),
\end{align*}
respectively, where $\hat{\theta}_{100(1-\alpha),i}$ represent the $100(1-\alpha)$\% posterior quantiles of $\theta_i$ and $\theta_i^*$ are true quantiles of $y$ at location $x_i$. 
Additional simulation results under a different true function are provided in the Supplementary Materials.

\subsection{Simulation results}
\label{subsec:4.2}

We show the simulation results for each scenario. 
Note that the point estimates of the variational Bayes method were the same as those of the calibrated variational method because the difference between them was only the variance of the variational posterior distribution. Hence, we omitted the results of the CVB-HS and CVB-Lap in Tables \ref{PC_table_MSEMAD} and \ref{VS_table_MSEMAD}. The frequentist quantile trend filtering by \cite{brantley2020baseline} is denoted by ``ADMM".

\begin{figure}[thbp]
\begin{center}
\includegraphics[width=\linewidth]{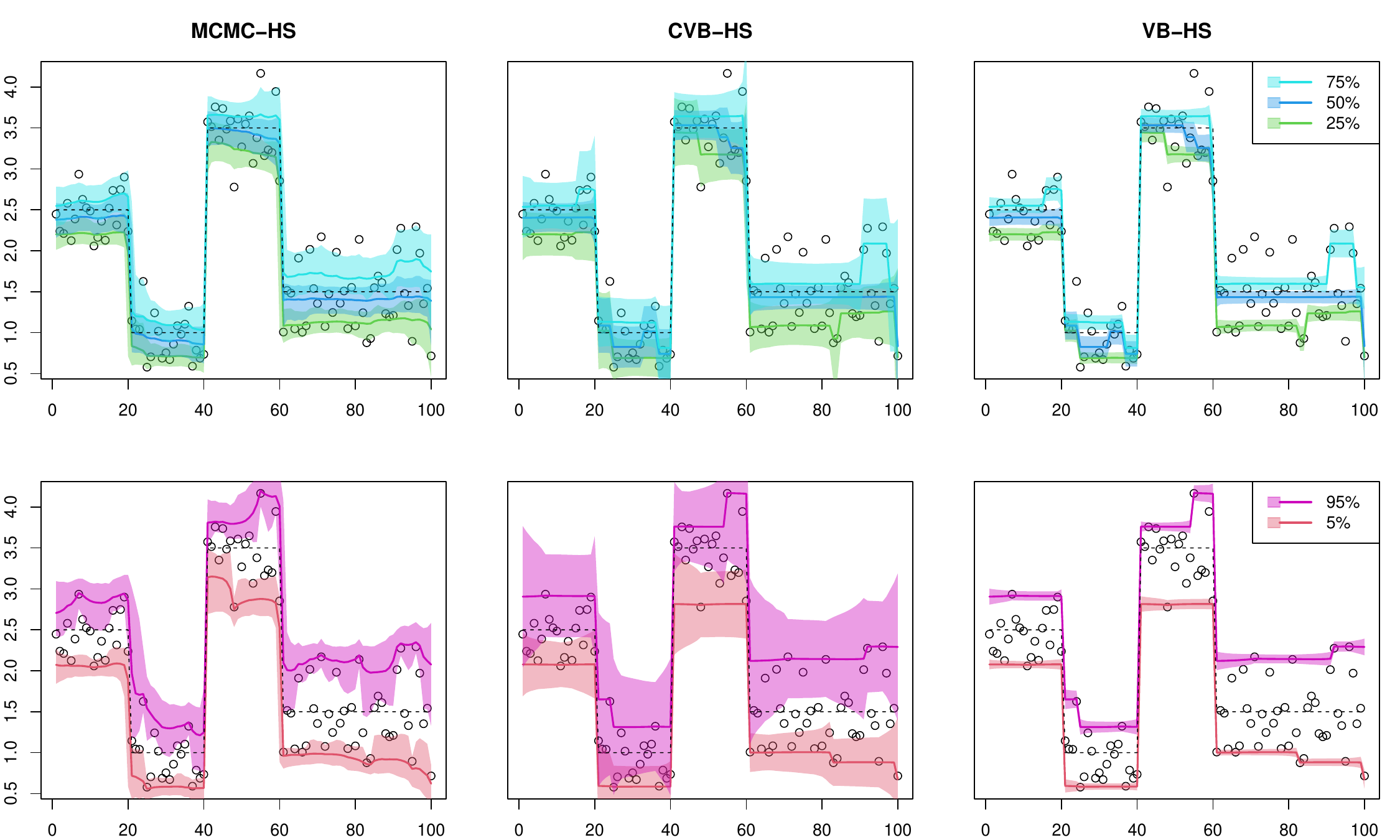}
\caption{One-shot simulation results under piecewise constant and Gauss noise. The order of trend filtering is $k=0$ for all methods.}
\label{PC_Gauss_0_6plot}
\end{center}
\end{figure}

\medskip
\noindent
{\it Piecewise constant}. We summarized the numerical results of the point estimate and uncertainty quantification in Tables \ref{PC_table_MSEMAD} and \ref{PC_table_MCIWCP}, respectively. From Table \ref{PC_table_MSEMAD}, we observed that the point estimates of the MCMC-HS method performed the best in all cases, and the frequentist ADMM method performed the worst in terms of MSE and MAD. 
For uncertainty quantification, it was shown that the MCMC methods have reasonable coverage probabilities for center quantiles such as $0.25$, $0.5$, and $0.75$ except for the case of beta distributed noise, while the MCMC methods for extremal quantiles such as $0.05$ and $0.95$ appear to be far away from the nominal coverage rate $0.95$. The MCIW of the VB-HS and VB-Lap methods tended to be shorter than that of the MCMC therefore, the corresponding coverage probabilities were extremely underestimated. However, the CVB-HS and CVB-Lap methods could quantify the uncertainty in almost all cases including extremal quantiles. We also show one-shot simulation results under the Gaussian noise in Figure \ref{PC_Gauss_0_6plot}. As shown in the figure, the credible intervals of CVB-HS are similar to those of MCMC-HS for $0.25$, $0.50$, and $0.75$ quantiles. Furthermore, the calibrated credible intervals by CVB-HS are wider than those of the MCMC for extremal quantiles.

\begin{table}[thbp]
\caption{Average values of MSE and MAD based on $100$ replications for piecewise constant with $k=0$.
The minimum values and second smallest values of MSE and MAD are represented in bold and italics respectively. }
\begin{center}
\resizebox{1.0\textwidth}{!}{ 
\begin{tabular}{c|ccccc|ccccc}
\toprule
& \multicolumn{5}{c|}{MSE} &  \multicolumn{5}{c}{MAD } \\
\hline
(I) Gauss & 0.05 & 0.25 & 0.5 & 0.75 & 0.95 & 0.05 & 0.25 & 0.5 & 0.75 & 0.95 \\ 
  \hline
MCMC-HS & {\bf 0.046} & {\bf 0.013} & {\bf 0.009} & {\bf 0.013} & {\bf 0.046} & {\bf 0.168} & {\bf 0.083} & {\bf 0.069} & {\bf 0.083} & {\bf 0.168} \\ 
  VB-HS &  0.094 & {\sl 0.033 } & {\sl 0.026 } & {\sl 0.034 } & {\sl 0.053 } & {\sl 0.198 } & {\sl 0.133 } & {\sl 0.112 } & {\sl 0.132 } & {\sl 0.182 } \\ 
  MCMC-Lap & 0.105 & 0.050 & 0.040 & 0.050 & 0.106 & 0.263 & 0.173 & 0.154 & 0.172 & 0.266 \\ 
  VB-Lap & {\sl 0.091} & 0.042 & 0.033 & 0.038 & 0.059 & 0.213 & 0.155 & 0.139 & 0.149 & 0.191 \\ 
  ADMM & 0.384 & 0.040 & 0.045 & 0.102 & 0.304 & 0.357 & 0.147 & 0.168 & 0.190 & 0.343 \\ 
\midrule
\midrule
(II) Beta & 0.05 & 0.25 & 0.5 & 0.75 &0.95 & 0.05 & 0.25 & 0.5 & 0.75 &0.95  \\
\hline
MCMC-HS & {\sl 0.004} & {\bf 0.003 } & {\bf 0.004} & {\bf 0.007 } & {\sl 0.020} & {\sl 0.026} & {\bf 0.027} & {\bf 0.040} & {\bf 0.056} & {\sl 0.109} \\ 
  VB-HS & {\bf 0.001} & {\sl 0.006} & {\sl 0.009} & {\sl 0.014} & {\sl 0.020} & {\bf 0.014} & {\sl 0.039} & {\sl 0.058} & {\sl 0.082} & 0.112 \\ 
  MCMC-Lap & 0.013 & 0.015 & 0.015 &  0.019 & 0.038 & 0.078 & 0.081 & 0.086 & 0.102 & 0.156 \\ 
  VB-Lap & 0.059 & 0.009 & 0.011 & {\sl 0.014} & {\bf 0.019 } & 0.083 & 0.060 & 0.072 & 0.086 & {\bf 0.106} \\ 
  ADMM & 0.793 & {\bf 0.003} & 0.011 & 0.101 & 0.442 & 0.437 & 0.041 & 0.085 & 0.164 & 0.411 \\ 
 \midrule
\midrule
(III) {\small Mixed normal} & 0.05 & 0.25 & 0.5 & 0.75 &0.95 & 0.05 & 0.25 & 0.5 & 0.75 &0.95  \\
\hline
MCMC-HS & {\bf 0.089} & {\bf 0.036} & {\bf 0.029} & {\bf 0.039} & {\bf 0.102} & {\bf 0.235} & {\bf 0.136} & {\bf 0.123} & {\bf 0.141} & {\bf 0.250} \\ 
  VB-HS & 0.247 & 0.080 & {\sl 0.067} & 0.085 & {\sl 0.125} & 0.316 & {\sl 0.209} & {\sl 0.190} & 0.216 & 0.275 \\ 
  MCMC-Lap & 0.191 & 0.094 & 0.078 & 0.095 & 0.204 & 0.365 & 0.240 & 0.217 & 0.242 & 0.379 \\ 
  VB-Lap & {\sl 0.137} & {\sl 0.079} & 0.069 & {\sl 0.079} & 0.126 & {\sl 0.282} & 0.213 & 0.201 & {\sl 0.215} & {\sl 0.273} \\ 
  ADMM & 0.315 & 0.183 & 0.123 & 0.170 & 0.236 & 0.386 & 0.271 & 0.266 & 0.284 & 0.351 \\ 
\bottomrule
\end{tabular}}
\end{center}
\label{PC_table_MSEMAD}
\end{table}

\begin{table}[thbp]
\caption{Average values of MCIW and CP based on $100$ replications for piecewise constant with $k=0$. The CP values above 90\% are represented in bold. }
\begin{center}
\resizebox{1.0\textwidth}{!}{ 
\begin{tabular}{c|ccccc|ccccc}
\toprule
& \multicolumn{5}{c|}{MCIW} &  \multicolumn{5}{c}{CP} \\
\hline
(I) Gauss & 0.05 & 0.25 & 0.5 & 0.75 & 0.95 & 0.05 & 0.25 & 0.5 & 0.75 & 0.95 \\ 
  \hline
MCMC-HS & 0.549 & 0.447 & 0.412 & 0.437 & 0.543 
& 0.757 & {\bf 0.950} & {\bf0.970} & {\bf0.943} & 0.761 \\ 
  CVB-HS & 1.034 & 0.620 & 0.471 & 0.596 & 1.015 
 & {\bf0.929} & {\bf0.924} & {\bf0.904} & {\bf0.920} & {\bf0.922} \\ 
   VB-HS & 0.117 & 0.205 & 0.221 & 0.205 & 0.117 
  & 0.189 & 0.477 & 0.611 & 0.462 & 0.187 \\
  MCMC-Lap & 0.775 & 0.764 & 0.785 & 0.759 & 0.762 
  & 0.806 & {\bf0.923} & {\bf0.960} & {\bf0.926} & 0.800 \\
  CVB-Lap & 1.039 & 0.822 & 0.606 & 0.792 & 0.990 
  & {\bf0.926} & {\bf0.952} & {\bf0.910} & {\bf0.953} & {\bf0.941} \\ 
  VB-Lap & 0.404 & 0.562 & 0.596 & 0.556 & 0.408 
  & 0.568 & 0.852 & {\bf0.907} & 0.871 & 0.607 \\
\midrule
\midrule
(II) Beta & 0.05 & 0.25 & 0.5 & 0.75 &0.95 & 0.05 & 0.25 & 0.5 & 0.75 &0.95  \\
\hline
MCMC-HS & 0.113 & 0.137 & 0.186 & 0.245 & 0.295 
& {\bf0.975} & {\bf0.959} & {\bf0.940} & {\bf0.916} & 0.676 \\ 
  CVB-HS & 0.282 & 0.202 & 0.207 & 0.360 & 0.640 
  & {\bf0.992} & {\bf0.938} & 0.864 & 0.894 & {\bf0.926} \\ 
  VB-HS & 0.034 & 0.069 & 0.094 & 0.099 & 0.063 
  & 0.805 & 0.682 & 0.573 & 0.405 & 0.167 \\
  MCMC-Lap & 0.377 & 0.326 & 0.326 & 0.309 & 0.390 
  & {\bf0.952} & {\bf0.927} & {\bf0.902} & 0.807 & 0.756 \\ 
  CVB-Lap & 0.281 & 0.305 & 0.307 & 0.385 & 0.407 
  & {\bf0.908} & {\bf0.929} & 0.899 & 0.889 & 0.835 \\ 
  VB-Lap & 0.214 & 0.274 & 0.307 & 0.296 & 0.264 
  & 0.855 & 0.916 & 0.899 & 0.827 & 0.676 \\
 \midrule
\midrule
(III) {\small Mixed normal} & 0.05 & 0.25 & 0.5 & 0.75 &0.95 & 0.05 & 0.25 & 0.5 & 0.75 &0.95  \\
\hline
MCMC-HS & 0.871 & 0.730 & 0.696 & 0.730 & 0.863 
& 0.797 & {\bf0.943} & {\bf0.972} & {\bf0.947} & 0.784 \\ 
  CVB-HS & 1.536 & 0.926 & 0.780 & 0.942 & 1.577 
  & {\bf0.929} & {\bf0.931} & {\bf0.919} & {\bf0.928} & {\bf0.955} \\ 
  VB-HS & 0.177 & 0.318 & 0.342 & 0.316 & 0.177 
  & 0.212 & 0.453 & 0.518 & 0.434 & 0.206 \\
  MCMC-Lap & 1.129 & 1.135 & 1.135 & 1.125 & 1.097 
  & 0.830 & {\bf0.940} & {\bf0.963} & {\bf0.942} & 0.795 \\ 
  CVB-Lap & 1.573 & 1.155 & 0.891 & 1.159 & 1.572 
  & {\bf0.947} & {\bf0.960} & {\bf0.920} & {\bf0.962} & {\bf0.954} \\ 
  VB-Lap & 0.548 & 0.778 & 0.818 & 0.770 & 0.556 
  & 0.575 & 0.852 & 0.897 & 0.851 & 0.595 \\
\bottomrule
\end{tabular}}
\end{center}
\label{PC_table_MCIWCP}
\end{table}

\begin{figure}[thbp]
\begin{center}
\includegraphics[width=\linewidth]{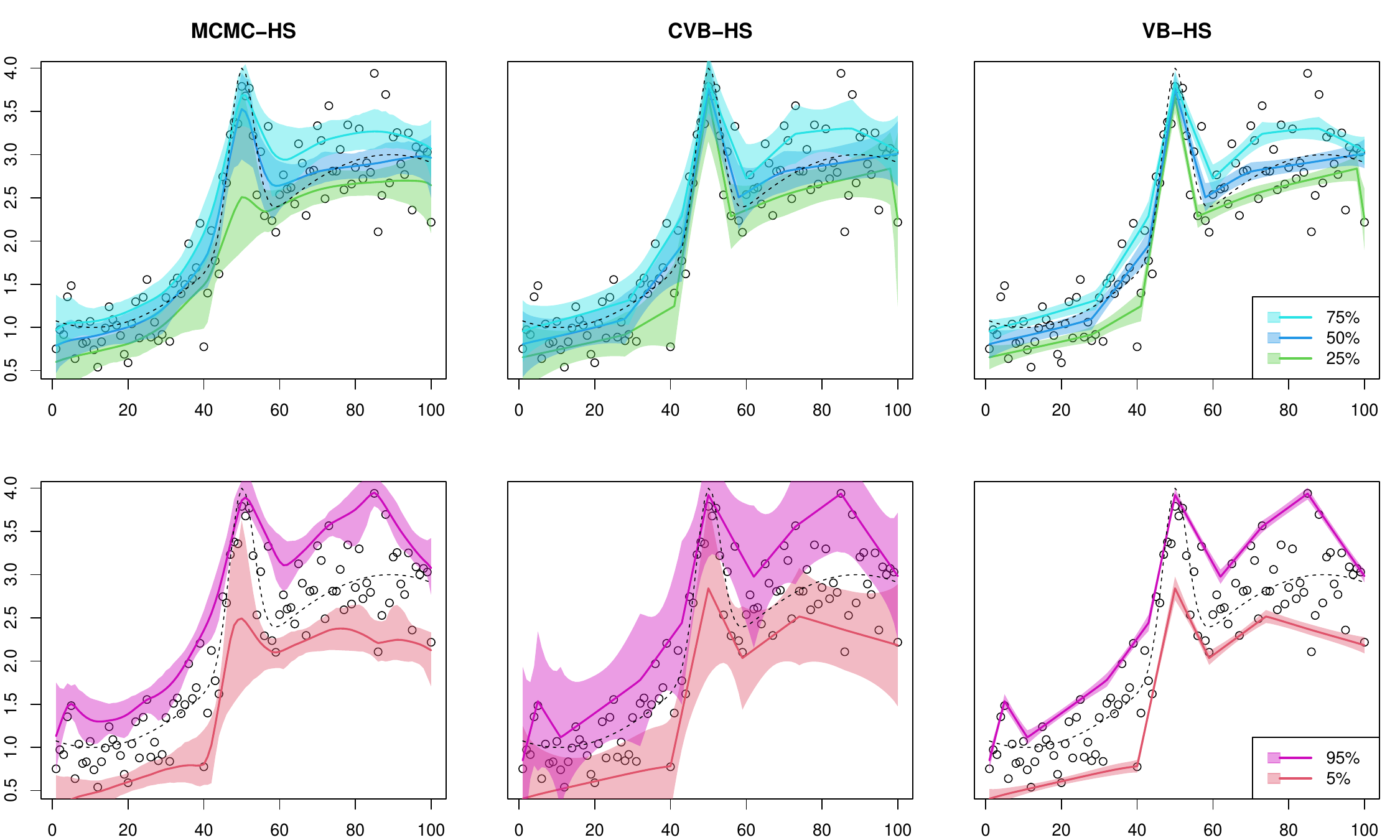}
\caption{One-shot simulation results under varying smoothness and Gauss noise. The order of trend filtering is $k=1$ for all methods.}
\label{VS_Gauss_1_6plot}
\end{center}
\end{figure}

\medskip
\noindent
{\it Varying smoothness}. The results for the (VS) scenario are reported in Tables \ref{VS_table_MSEMAD} and \ref{VS_table_MCIWCP}. 
From Table \ref{VS_table_MSEMAD}, the MCMC-HS method also performed well relative to the other methods in terms of point estimation, while the variational Bayes methods under horseshoe prior also provided comparable point estimates. Different from the (PC) scenario, the MCMC methods had slightly worse coverage probabilities. In particular, the MCMC-HS and MCMC-Lap under the mixed normal noise, which is a relatively high degree of misspecification, appeard to be far from the nominal coverage rate of $0.95$. Although the MCIW of the variational Bayes methods without calibration also tended to be shorter than that of MCMC, the calibrated variational Bayes dramatically improved the coverage even under the mixed normal case. We also show one-shot simulation results under the Gaussian noise in Figure \ref{VS_Gauss_1_6plot}.

Finally, we assessed the efficiency of posterior computation. 
To this end, we calculated the raw computing time and effective sample size per unit time. The latter is defined as the effective sample size divided by the computation time in seconds.
Note that the effective sample size for the variational Bayes methods (VB and CVB) was 7,500 since i.i.d. samples could be drawn from variational posterior distributions.
The values averaged over 100 replications of simulating datasets are presented in Table~\ref{table_timeESS}. 
The results show that the proposed algorithm provides posterior samples much more efficiently than the MCMC algorithm. 
Such computationally efficient property of the proposed method is a benefit of a novel combination of variational approximation and posterior calibration. 

\begin{table}[thbp]
\caption{Average values of MSE and MAD based on $100$ replications for varying smoothness with $k=1$.
The minimum values and second smallest values of MSE and MAD are represented in bold and italics respectively.}
\begin{center}
\resizebox{1.0\textwidth}{!}{ 
\begin{tabular}{c|ccccc|ccccc}
\toprule
& \multicolumn{5}{c|}{MSE} &  \multicolumn{5}{c}{MAD } \\
\hline
(I) Gauss & 0.05 & 0.25 & 0.5 & 0.75 & 0.95 & 0.05 & 0.25 & 0.5 & 0.75 & 0.95 \\ 
  \hline
MCMC-HS & {\sl 0.068} & {\sl 0.034} & {\bf 0.017} & {\bf 0.019} & {\bf 0.041} & {\bf 0.179} & {\bf 0.119} & {\bf 0.097} & {\bf 0.104} & {\bf 0.156} \\ 
  VB-HS & 0.133 & {\bf 0.026} & {\sl 0.020} & {\sl 0.025} & {\sl 0.056} & 0.229 & {\bf 0.119} & {\sl 0.105} & {\sl 0.117} & {\sl 0.180} \\ 
  MCMC-Lap & {\bf 0.064} & 0.039 & 0.026 & 0.028 & 0.057 & {\sl 0.194} & {\sl 0.138} & 0.122 & 0.128 & 0.188 \\ 
  VB-Lap & 0.097 & 0.052 & 0.027 & 0.028 & 0.058 & 0.209 & 0.143 & 0.121 & 0.128 & 0.190 \\ 
  ADMM & 0.177 & 0.075 & 0.031 & 0.035 & 0.097 & 0.237 & 0.163 & 0.123 & 0.137 & 0.230 \\ 
\midrule
\midrule
(II) Beta & 0.05 & 0.25 & 0.5 & 0.75 &0.95 & 0.05 & 0.25 & 0.5 & 0.75 &0.95  \\
\hline
MCMC-HS & {\bf 0.004} & {\bf 0.004} & {\bf 0.004} & {\bf 0.007} & {\bf 0.014} & {\bf 0.042} & {\bf 0.037} & {\bf 0.046} & {\bf 0.058} & {\bf 0.092} \\ 
  VB-HS & 0.101 & {\sl 0.005} & {\sl 0.006} & {\sl 0.009} & {\sl 0.020} & 0.107 & 0.044 & {\sl 0.053} & {\sl 0.069} & {\sl 0.106} \\ 
  MCMC-Lap & {\sl 0.006} & {\sl 0.005} & {\sl 0.006} & {\sl 0.009} & 0.022 & {\sl 0.054} & 0.046 & 0.057 & 0.072 & 0.115 \\ 
  VB-Lap & 0.022 & {\sl 0.005} & {\sl 0.006} & 0.010 & 0.021 & 0.067 & {\sl 0.043} & 0.056 & 0.072 & 0.113 \\ 
  ADMM & 0.204 & 0.057 & 0.011 & 0.018 & 0.094 & 0.171 & 0.100 & 0.068 & 0.088 & 0.199 \\ 
 \midrule
\midrule
(III) {\small Mixed normal} & 0.05 & 0.25 & 0.5 & 0.75 &0.95 & 0.05 & 0.25 & 0.5 & 0.75 &0.95  \\
\hline
MCMC-HS & {\sl 0.147} & 0.130 & 0.077 & {\bf 0.055} & {\bf 0.090} & {\bf 0.253} & 0.214 & 0.181 & {\bf 0.172} & {\bf 0.230} \\ 
  VB-HS & 0.167 & {\bf 0.062} & {\bf 0.047} & {\sl 0.056} & 0.114 & 0.279 & {\bf 0.181} & {\bf 0.162} & {\sl 0.178} & {\sl 0.258} \\ 
  MCMC-Lap & {\bf 0.108} & {\sl 0.085} & {\sl 0.061} & 0.060 & {\sl 0.112} & {\sl 0.255} & {\sl 0.198} & {\sl 0.180} & 0.188 & 0.265 \\ 
  VB-Lap & 0.147 & 0.106 & 0.072 & 0.066 & 0.114 & 0.274 & 0.206 & 0.186 & 0.194 & 0.263 \\ 
  ADMM & 0.195 & 0.088 & 0.057 & 0.066 & 0.138 & 0.284 & 0.202 & 0.174 & 0.192 & 0.287 \\ 
\bottomrule
\end{tabular}}
\end{center}
\label{VS_table_MSEMAD}
\end{table}

\begin{table}[thbp]
\caption{Average values of MCIW and CP based on $100$ replications for varying smoothness with $k=1$. The CP values above 90\% are represented in bold. }
\begin{center}
\resizebox{1.0\textwidth}{!}{ 
\begin{tabular}{c|ccccc|ccccc}
\toprule
& \multicolumn{5}{c|}{MCIW} &  \multicolumn{5}{c}{CP} \\
\hline
(I) Gauss & 0.05 & 0.25 & 0.5 & 0.75 & 0.95 & 0.05 & 0.25 & 0.5 & 0.75 & 0.95 \\ 
  \hline
MCMC-HS & 0.498 & 0.476 & 0.447 & 0.444 & 0.446 & 0.730 & 0.890 & {\bf 0.930} & {\bf 0.903} & 0.728 \\ 
  CVB-HS & 1.010 & 0.592 & 0.496 & 0.572 & 1.097 & {\bf 0.911} & {\bf 0.935} & {\bf 0.936} & {\bf 0.937} & {\bf 0.956} \\ 
  VB-HS & 0.124 & 0.196 & 0.209 & 0.195 & 0.118 & 0.206 & 0.505 & 0.584 & 0.518 & 0.215 \\ 
  MCMC-Lap & 0.545 & 0.571 & 0.562 & 0.553 & 0.524 & 0.760 & 0.897 & {\bf 0.929} & {\bf 0.914} & 0.756 \\ 
  CVB-Lap & 1.335 & 0.890 & 0.602 & 0.821 & 1.240 & {\bf 0.961} & {\bf 0.960} & {\bf 0.937} & {\bf 0.977} & {\bf 0.968} \\ 
  VB-Lap & 0.207 & 0.326 & 0.369 & 0.348 & 0.227 & 0.346 & 0.705 & 0.788 & 0.726 & 0.364 \\ 
\midrule
\midrule
(II) Beta & 0.05 & 0.25 & 0.5 & 0.75 &0.95 & 0.05 & 0.25 & 0.5 & 0.75 &0.95  \\
\hline
MCMC-HS & 0.162 & 0.178 & 0.213 & 0.241 & 0.260 & {\bf 0.931} & {\bf 0.941} & {\bf 0.925} & 0.898 & 0.713 \\ 
  CVB-HS & 0.445 & 0.259 & 0.209 & 0.347 & 0.724 & {\bf 0.927} & {\bf 0.950} & 0.884 & {\bf 0.935} & {\bf 0.969} \\ 
  VB-HS & 0.051 & 0.085 & 0.103 & 0.104 & 0.065 & 0.502 & 0.656 & 0.606 & 0.487 & 0.219 \\ 
  MCMC-Lap & 0.251 & 0.248 & 0.282 & 0.306 & 0.304 & 0.953 & 0.952 & 0.942 & 0.903 & 0.736 \\ 
  CVB-Lap & 0.522 & 0.362 & 0.268 & 0.458 & 0.673 & {\bf 0.968} & {\bf 0.976} & {\bf 0.923} & {\bf 0.969} & {\bf 0.954} \\ 
  VB-Lap & 0.108 & 0.164 & 0.199 & 0.201 & 0.136 & 0.723 & 0.874 & 0.854 & 0.756 & 0.368 \\ 
 \midrule
\midrule
(III) {\small Mixed normal} & 0.05 & 0.25 & 0.5 & 0.75 &0.95 & 0.05 & 0.25 & 0.5 & 0.75 &0.95  \\
\hline
MCMC-HS & 0.600 & 0.636 & 0.654 & 0.655 & 0.666 & 0.692 & 0.838 & 0.874 & 0.876 & 0.743 \\ 
  CVB-HS & 1.271 & 0.854 & 0.747 & 0.850 & 1.540 & {\bf 0.909} & {\bf 0.926} & {\bf 0.927} & {\bf 0.934} & {\bf 0.961} \\ 
  VB-HS & 0.172 & 0.286 & 0.303 & 0.284 & 0.178 & 0.211 & 0.490 & 0.558 & 0.496 & 0.226 \\
  MCMC-Lap & 0.727 & 0.755 & 0.765 & 0.768 & 0.755 & 0.753 & 0.885 & {\bf 0.907} & 0.891 & 0.749 \\ 
  CVB-Lap & 1.780 & 1.214 & 0.891 & 1.249 & 1.972 & {\bf 0.963} & {\bf 0.942} & {\bf 0.924} & {\bf 0.969} & {\bf 0.979} \\ 
  VB-Lap & 0.279 & 0.400 & 0.458 & 0.455 & 0.324 & 0.330 & 0.647 & 0.711 & 0.677 & 0.394 \\ 
\bottomrule
\end{tabular}}
\end{center}
\label{VS_table_MCIWCP}
\end{table}

\begin{table}[thbp]
\caption{Average values of raw computing time and effective sample size per unit time based on $100$ replications for all scenarios.}
\begin{center}
\resizebox{1.0\textwidth}{!}{ 
\begin{tabular}{c|ccccc|ccccc}
\toprule
\multicolumn{11}{c}{(PC) Piecewise constant}\\
\midrule
& \multicolumn{5}{c|}{Computation time (second)} &  \multicolumn{5}{c}{ESS (per second)} \\
\midrule
(I) Gauss & 0.05 & 0.25 & 0.5 & 0.75 & 0.95 & 0.05 & 0.25 & 0.5 & 0.75 & 0.95 \\ 
  \hline
MCMC-HS & 33 & 32 & 32 & 32 & 33 & 13 & 39 & 45 & 40 & 13 \\ 
  CVB-HS & 13 & 9 & 11 & 9 & 12 & 603 & 869 & 699 & 867 & 613 \\ 
  MCMC-Lap & 37 & 36 & 36 & 36 & 37 & 12 & 81 & 109 & 82 & 13 \\ 
  CVB-Lap & 12 & 9 & 9 & 7 & 11 & 662 & 848 & 862 & 1120 & 687 \\ 
\hline
\hline
(II) Beta & 0.05 & 0.25 & 0.5 & 0.75 &0.95 & 0.05 & 0.25 & 0.5 & 0.75 &0.95  \\
\hline
MCMC-HS & 32 & 32 & 32 & 32 & 33 & 13 & 61 & 59 & 41 & 13 \\ 
  CVB-HS & 8 & 7 & 9 & 8 & 10 & 1006 & 1089 & 874 & 991 & 792 \\ 
  MCMC-Lap & 37 & 37 & 37 & 37 & 37 & 11 & 79 & 126 & 74 & 12 \\ 
  CVB-Lap & 6 & 5 & 8 & 4 & 6 & 1306 & 1654 & 894 & 1791 & 1287 \\ 
\hline
\hline
(III) {\small Mixed normal} & 0.05 & 0.25 & 0.5 & 0.75 &0.95 & 0.05 & 0.25 & 0.5 & 0.75 &0.95  \\
\hline
MCMC-HS & 32 & 31 & 31 & 32 & 32 & 13 & 35 & 40 & 35 & 13 \\ 
  CVB-HS & 16 & 11 & 12 & 10 & 15 & 493 & 724 & 606 & 731 & 497 \\ 
  MCMC-Lap & 36 & 36 & 36 & 36 & 36 & 13 & 78 & 98 & 79 & 14 \\ 
  CVB-Lap & 15 & 12 & 9 & 9 & 13 & 513 & 634 & 807 & 883 & 586 \\
\midrule
\midrule
\multicolumn{11}{c}{(VS) Varying smoothness}\\
\midrule
& \multicolumn{5}{c|}{Computation time (second)} &  \multicolumn{5}{c}{ESS (per second)} \\
\midrule
(I) Gauss & 0.05 & 0.25 & 0.5 & 0.75 & 0.95 & 0.05 & 0.25 & 0.5 & 0.75 & 0.95 \\ 
  \hline
MCMC-HS & 32 & 32 & 32 & 32 & 32 & 13 & 31 & 37 & 35 & 14 \\ 
  CVB-HS & 12 & 8 & 11 & 8 & 13 & 620 & 897 & 679 & 923 & 575 \\ 
  MCMC-Lap & 36 & 35 & 35 & 35 & 36 & 13 & 53 & 72 & 65 & 13 \\ 
  CVB-Lap & 16 & 13 & 11 & 11 & 15 & 471 & 588 & 684 & 713 & 519 \\
\hline
\hline
(II) Beta & 0.05 & 0.25 & 0.5 & 0.75 &0.95 & 0.05 & 0.25 & 0.5 & 0.75 &0.95  \\
\hline
MCMC-HS & 33 & 33 & 33 & 33 & 33 & 12 & 48 & 45 & 38 & 14 \\ 
  CVB-HS & 9 & 7 & 10 & 7 & 11 & 861 & 1118 & 780 & 1075 & 699 \\ 
  MCMC-Lap & 36 & 36 & 36 & 36 & 36 & 11 & 78 & 105 & 69 & 12 \\ 
  CVB-Lap & 11 & 6 & 9 & 6 & 9 & 740 & 1202 & 813 & 1181 & 879 \\ 
\hline
\hline
(III) {\small Mixed normal} & 0.05 & 0.25 & 0.5 & 0.75 &0.95 & 0.05 & 0.25 & 0.5 & 0.75 &0.95  \\
\hline
MCMC-HS & 32 & 32 & 32 & 32 & 32 & 14 & 25 & 30 & 30 & 15 \\ 
  CVB-HS & 14 & 10 & 13 & 10 & 16 & 555 & 756 & 593 & 779 & 481 \\ 
  MCMC-Lap & 36 & 36 & 36 & 36 & 36 & 14 & 36 & 49 & 51 & 14 \\ 
  CVB-Lap & 22 & 20 & 17 & 19 & 23 & 359 & 392 & 463 & 412 & 335 \\ 
\bottomrule
\end{tabular}}
\end{center}
\label{table_timeESS}
\end{table}

\section{Real data analysis}
\label{sec:5}

\subsection{Nile data}
\label{subsec:5.1}

\begin{figure}[htbp]
\begin{center}
\includegraphics[width=\linewidth]{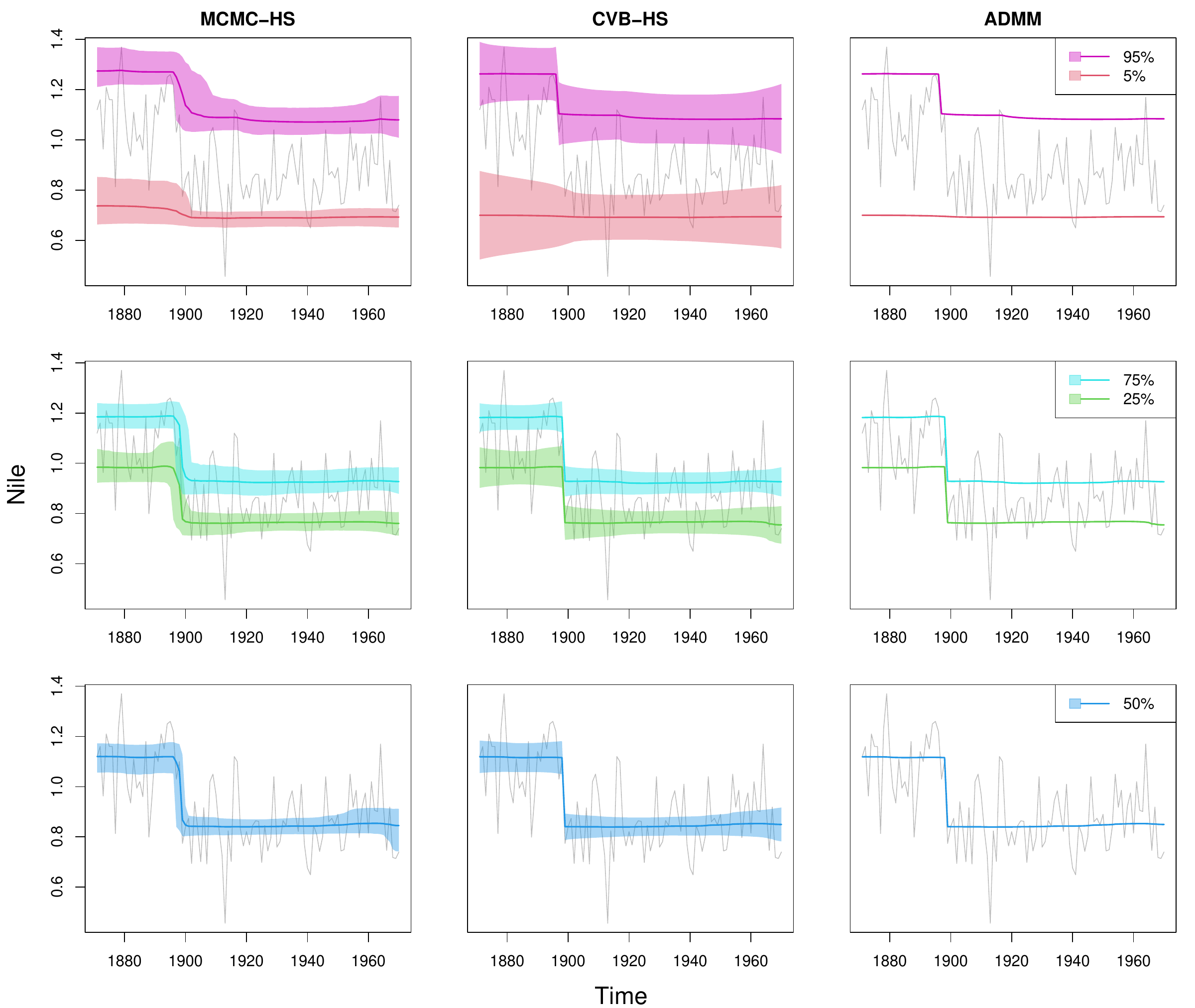}
\caption{Point estimates and 95\% credible intervals for Nile data. }
\label{Nile_plot}
\end{center}
\end{figure}

We first applied the proposed methods to the famous Nile river data \citep{cobb1978problem, balke1993detecting}. The data contains measurements of the annual flow of the river Nile from 1871 to 1970, and we found an apparent change-point near 1898. We considered $k=0$ and compared the three methods, that is MCMC-HS, CVB-HS, and ADMM. We generated 60,000 posterior samples after discarding the first 10,000 posterior samples as burn-in, and then only every 10th scan was saved. For the Bayesian methods, we adopted $(a,b)=(1,3)$ as hyperparameters in the inverse gamma prior to $\sigma^2$. The resulting estimates of quantiles and the corresponding 95\% credible intervals are shown in Figure \ref{Nile_plot}. In terms of point estimation, the horseshoe prior appears to capture the piecewise constant structures well, and the point estimates of CVB-HS and ADMM are comparable for all quantiles. For uncertainty quantification, the lengths of credible intervals of the MCMC-HS and CVB-HS are comparable for 25\%, 50\%, and 75\% quantiles, while the CVB-HS method has wider credible intervals than those of the MCMC method especially for extremal quantiles such as 5\% and 95\% (see also Table \ref{realdata_table_AL}). This is consistent with the simulation results in Section \ref{subsec:4.2}. 
In Table~\ref{realdata_table_ESS}, we provided the effective sample size per unit time of the proposed algorithm and MCMC, which showed significant improvement of computational efficiency by the proposed method.
Hence, we could conclude that the proposed algorithm performs better than the MCMC for this application, in terms of both qualities of inference and computational efficiency.

\begin{table}[thbp]
\caption{Average lengths of credible intervals for real data examples.}
\begin{center}
\resizebox{1.0\textwidth}{!}{ 
\begin{tabular}{c|ccccc|ccccc}
\toprule
& \multicolumn{5}{c|}{Nile data (Section \ref{subsec:5.1})} &  \multicolumn{5}{c}{Munich rent data (Section \ref{subsec:5.2})} \\
\hline
 & 0.05 & 0.25 & 0.5 & 0.75 & 0.95 & 0.1 & 0.3 & 0.5 & 0.7 & 0.9 \\ 
  \hline
MCMC-HS & 0.10 & 0.10 & 0.10 & 0.11 & 0.14 & 0.98 & 0.88 & 0.93 & 0.90 & 0.95\\ 
  CVB-HS & 0.23 & 0.12 & 0.10 & 0.10 & 0.21 & 2.33 & 1.37 & 1.15 & 1.30 & 1.88 \\ 
\bottomrule
\end{tabular}}
\end{center}
\label{realdata_table_AL}
\end{table}

\begin{table}[thbp]
\caption{Effective sample size per unit time for real data examples.}
\begin{center}
\resizebox{1.0\textwidth}{!}{ 
\begin{tabular}{c|ccccc|ccccc}
\toprule
& \multicolumn{5}{c|}{Nile data (Section \ref{subsec:5.1})} &  \multicolumn{5}{c}{Munich rent data (Section \ref{subsec:5.2})} \\
\hline
& 0.05 & 0.25 & 0.5 & 0.75 & 0.95 & 0.1 & 0.3 & 0.5 & 0.7 & 0.9 \\
  \hline
MCMC-HS & 4 & 7 & 7 & 6 & 5 & 4 & 7 & 7 & 6 & 5 \\ 
  CVB-HS & 47619 & 52632 & 47619 & 55556 & 48780 & 4029 & 6233 & 1943 & 6124 & 2949\\
\bottomrule
\end{tabular}}
\end{center}
\label{realdata_table_ESS}
\end{table}

\subsection{Munich rent data}\label{subsec:5.2}

The proposed methods can also be applied to multiple observations with an irregular grid. We used Munich rent data (\url{https://github.com/jrfaulkner/spmrf}) which includes the value of rent per square meter and floor space in Munich, Germany \citep[see also][]{rue2005gaussian, faulkner2018locally, heng2022bayesian}. The data has multiple observations per location and an irregular grid. Let the response $y=(y_1,\dots,y_n)$ be the rent and the location $x=(x_1,\dots,x_n)$ be the floor size. At the  location $x_i$, the response $y_i$ has multiple observations per location, that is, $y_i=(y_{i1},\dots,y_{iN_i})^{\top} \in \mathbb{R}^{N_i}$. Furthermore, the difference $x_{j+1}-x_j$ is not always constant, therefore the floor spaces are unequally spaced. This is a different situation from the example in Section~\ref{subsec:5.1}. The data contains $N=\sum_{i=1}^nN_i=2,035$ observations and the floor space (or location) has 134 distinct values. We applied the third-order adjusted difference operator defined in Remark \ref{rem1} to the proposed Bayesian quantile trend filtering methods (i.e. MCMC-HS and CVB-HS with $k=2$). Since Brantley's quantile trend filtering method \citep{brantley2020baseline} cannot be applied to the data with multiple observations per location, we applied the quantile smoothing spline method by \cite{nychka2017fields} as a frequentist competitor. The method could be implemented by using \texttt{qsreg} function in {\texttt R} package  \texttt{fields}. The details of the method are provided in \cite{nychka1995nonparametric} and \cite{oh2004period}, and the smoothing parameter was chosen by using cross-validation. 
For these methods, we analyzed the five quantile levels such as $10\%$, $30\%$, $50\%$, $70\%$ and $90\%$. For the Bayesian methods, we generated 60,000 posterior samples after discarding the first 10,000 posterior samples as burn-in, and then only every 10th scan was saved. 

\begin{figure}[htbp]
\begin{center}
\includegraphics[width=\linewidth]{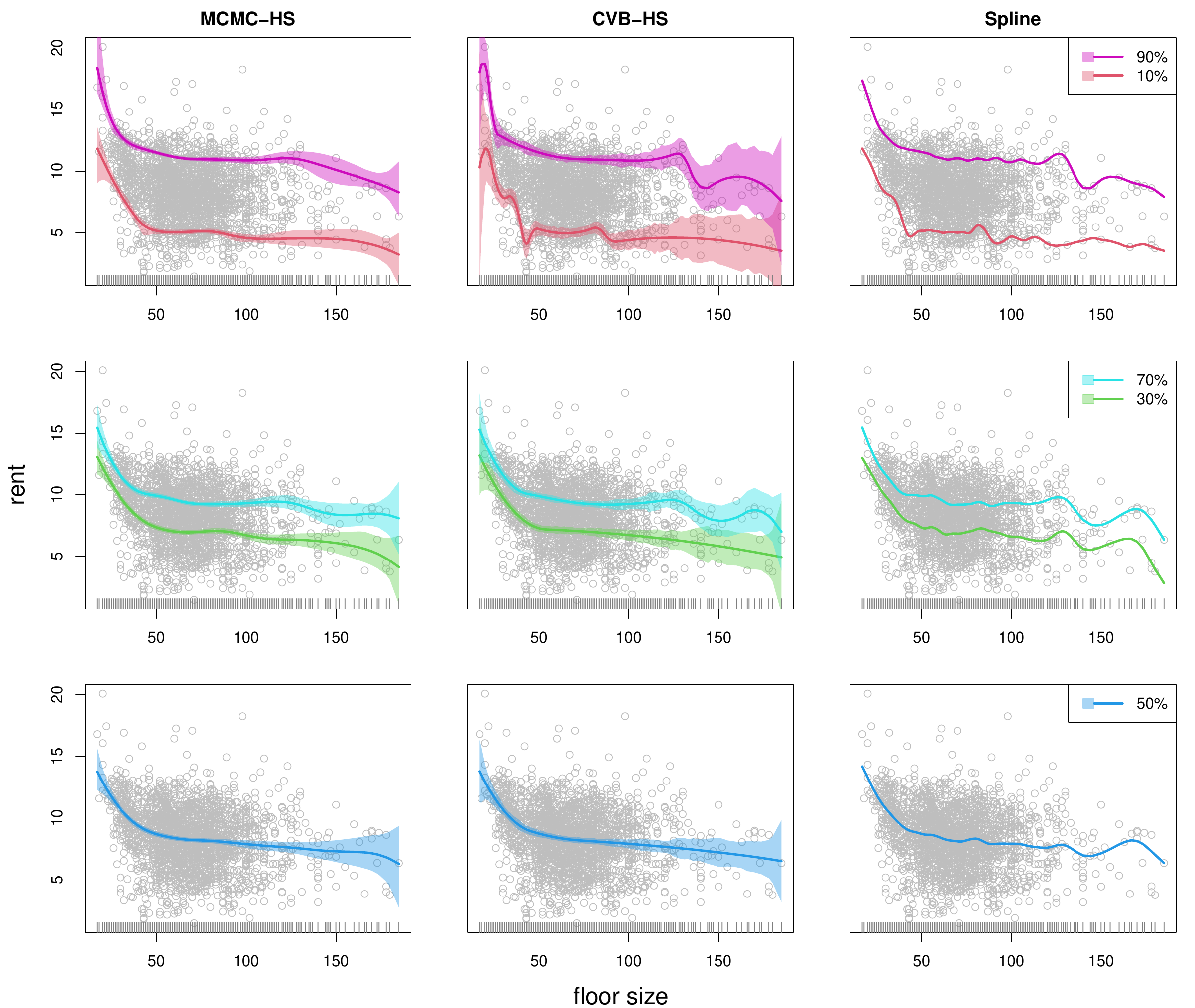}
\caption{Point estimates and 95\% credible intervals for Munich rent data.}
\label{munich_plot}
\end{center}
\end{figure}

The results of the point estimate and credible interval are shown in Figure \ref{munich_plot}. The frequentist smoothing spline method is denoted by ``Spline" in Figure \ref{munich_plot}.
The CVB-HS and Spline methods gave comparable baseline estimates, while the MCMC-HS method provided slightly smoother point estimates than the other two methods, especially for the large floor size. These decreasing trends mean that the houses with small floor sizes have a greater effect on their rent. Such a trend was also observed in the Bayesian mean trend filtering by \cite{faulkner2018locally}. Compared with the MCMC-HS, the CVB-HS method has a wider length of 95\% credible intervals for large values of floor size. The phenomenon appears to be reasonable because the data are less in such regions.
Additionally, two Bayesian methods provided almost the same results for the center quantile levels such as 30\%, 50\%, and 70\%, while the credible intervals of CVB-HS were wider than those of the MCMC-HS especially for extremal quantile levels such as 90\% and 10\% (see also Table \ref{realdata_table_AL}). This indicates that the MCMC-HS method possibly underestimated extremal quantile regions.
We again computed the effective sample size per unit time of the proposed algorithm and the MCMC, and the results are given in Table~\ref{realdata_table_ESS}.
From the results, we concluded that the proposed algorithm performs better than the MCMC for this application, in terms, not only of the quality of inference, but also of computational efficiency.

\section{Discussion}

\section{Concluding remarks}\label{sec:conc}
This study proposed a quick and accurate calibration algorithm for credible intervals using a mean-field variational Bayes method. The proposed CVB method can reasonably calibrate credible intervals with possible model misspecifications. In numerical experiments, it was shown that the proposed method worked especially well in the inference for high/low quantile levels. We also showed that the computational efficiency of the proposed CVB methods is higher than the MCMC versions in terms of the efficient sample size and computation time. If computation time is not a concern, then MCMC-based methods may be capable of providing accurate point and interval estimates. However, the estimation of low/high quantile tends to be unstable, and if the model is misspecified, estimation of other quantile points will also be unstable. The method of \cite{syring2019calibrating} could also be used in such cases, but the proposed CVB method is capable of parallel computation and is thus much more computationally efficient. Finally, as drawbacks of the variational posterior approximations, the proposed CVB method may not accurately reflect prior beliefs about parameters. It was observed that the CVB-HS and Spline had remarkably similar results in terms of the trajectories of the trends in the Munich rent example. We believe this is due to the variational approximation of the posterior distribution. However, the proposed method still has the advantage of providing point estimation results comparable to those of the optimization method and of allowing the quick and accurate evaluation of uncertainty under finite samples. 

In our study, we focused on the use of asymmetric Laplace likelihood, but it would be possible to extend the framework to extended asymmetric Laplace \citep[e.g.][]{yan2017new}. 
In the setting, it is possible to develop a similar computation algorithm to obtain posterior distribution and a valid credible interval, which will be left to future research. 
Furthermore, it may be more suitable to use a skewed distribution as a variational distribution of the quantile trends which can provide asymmetric credible intervals. 
However, when a different variational distribution was adopted, the mean-filed approximation algorithm used in this paper was no longer applicable, therefore a detailed investigation extends the scope of this study.

\section*{Acknowledgement}
The authors would like to thank the Associate Editor and a reviewer for their valuable comments and suggestions to improve the quality of this article. 
This work was supported by JST, the establishment of university fellowships towards the creation of science technology innovation, Grant Number JPMJFS2129. This work is partially supported by Japan Society for Promotion of Science (KAKENHI) grant numbers 21K13835 and 21H00699.

\vspace{5mm}
\bibliography{ref}
\bibliographystyle{chicago}

\newpage
\setcounter{page}{1}
\setcounter{equation}{0}
\renewcommand{\theequation}{S\arabic{equation}}
\setcounter{section}{0}
\renewcommand{\thesection}{S\arabic{section}}
\setcounter{table}{0}
\renewcommand{\thetable}{S\arabic{table}}
\setcounter{figure}{0}
\renewcommand{\thefigure}{S\arabic{figure}}

\begin{center}
{\LARGE\bf Supplementary Materials for ``Fast and Locally Adaptive Bayesian Quantile Smoothing 
using Calibrated Variational Approximations"}
\end{center}

\vspace{1cm}
This Supplementary Material provides algorithm details and additional information on simulation study and real data examples.


\section{Full conditional distributions in Gibbs sampler}

We provide the details of the full conditional distributions. 

\begin{itemize}
\item The full conditional distribution of $\theta$.
\begin{align*}
    p(\theta\mid \tau^2, w, z, \sigma^2,y)
    &\propto \prod_{i=1}^n\prod_{j=1}^{N_i}\exp\left(-\frac{(y_{ij}-\theta_i-\psi z_{ij})^2}{2t^2 \sigma^2z_{ij}}\right)\exp\left(-\frac{1}{2\sigma^2}\theta^{\top}D^{\top}W^{-1}D\theta\right)\\
    &\propto \exp\left\{-\frac{1}{2\sigma^2}(\theta-A^{-1}B)^{\top}A(\theta-A^{-1}B)\right\},
\end{align*}
which is $N_n(A^{-1}B, \sigma^2A^{-1})$, where 
\begin{align*}
    a_i&=\sum_{j=1}^{N_i}\frac{1}{ z_{ij}}, \quad b_i= \sum_{j=1}^{N_i}\frac{( y_{ij}-\psi z_{ij})}{ z_{ij}}, \quad c=\left(\sum_{j=1}^{N_1}y_{1j}-\psi z_{1j},\dots, \sum_{j=1}^{N_n}y_{nj}-\psi z_{nj}\right)^{\top}\\
    A&=D^{\top}W^{-1}D+\frac{1}{t^2}\diag\left(\sum_{j=1}^{N_1}z_{1j}^{-1},\ldots,\sum_{j=1}^{N_n}z_{nj}^{-1}\right), \\
    B&=\left(\frac{1}{t^2}\sum_{j=1}^{N_1}\left( \frac{y_{1j}}{z_{1j}}-\psi\right),\dots,\frac{1}{t^2}\sum_{j=1}^{N_n}\left( \frac{y_{nj}}{z_{nj}}-\psi\right)\right)^{\top}.
\end{align*}
\item The full conditional distributions of $z_{ij}$ for $i=1,\dots,n$ and $j=1,\dots, N_i$.
\begin{align*}
    p(z_{ij}\mid \theta_i, y, \sigma^2)
    & \propto (z_{ij})^{-1/2}\exp\left(-\frac{(y_{ij}-\theta_i-\psi z_{ij})^2}{2t^2\sigma^2 z_{ij}}\right) \exp\left(-\frac{z_{ij}}{\sigma^2}\right)\\
    &\propto (z_{ij})^{-1+1/2}\exp\left\{-\frac{1}{2}\left(\frac{ (y_{ij}-\theta_i)^2}{t^2\sigma^2}\frac{1}{z_{ij}} + \left(\frac{\psi^2}{t^2}+2\right)\frac{1}{\sigma^2}z_{ij}\right)\right\},
\end{align*}
which is $\mathrm{GIG}\left(1/2, (y_{ij}-\theta_i)^2/(t^2\sigma^2), \left(\psi^2/t^2+2\right)z_{ij}/\sigma^2 \right)$.
\item The full conditional distribution of $\sigma^2$.
\begin{align*}
    &p(\sigma^2\mid \theta,\tau^2,w,y,z)\\
    &\propto
    \prod_{i=1}^n\prod_{j=1}^{N_i}(\sigma^2)^{-1/2}\exp\left(-\frac{(y_{ij}- \theta_i-\psi z_{ij})^2}{2t^2 \sigma^2z_{ij}}\right)
    \times (\sigma^2)^{-n/2}\exp\left(-\frac{1}{2\sigma^2}\theta^{\top}D^{\top}W^{-1}D\theta\right)\\
    &\quad\quad \times \prod_{i=1}^n\prod_{j=1}^{N_i} (\sigma^2)^{-1}\exp\left(-\frac{z_{ij}}{\sigma^2}\right)
    \times (\sigma^2)^{-1-a_{\sigma}}\exp\left(-\frac{b_{\sigma}}{\sigma^2}\right)\\
    &\propto
    (\sigma^2)^{-1-(3N+n)/2+a_{\sigma}}\\
    &\quad \times \exp\left\{-\frac{1}{\sigma^2}\left(\sum_{i=1}^n\sum_{j=1}^{N_i}\frac{(y_{ij}-\theta_i-\psi z_{ij})^2}{2t^2 z_{ij}}+\frac{1}{2}\theta^{\top}D^{\top}W^{-1}D\theta+\sum_{i=1}^n\sum_{j=1}^{N_i}z_{ij}+b_{\sigma}\right)\right\},
\end{align*}
which is $\mathrm{IG}\left((3N+n)/2+a_{\sigma}, \alpha_{\sigma^2}\right)$, where
\begin{align*}
    \alpha_{\sigma^2}=\sum_{i=1}^n\sum_{j=1}^{N_i}\frac{(y_{ij}-\theta_i-\psi z_{ij})^2}{2t^2 z_{ij}}+\frac{1}{2}\theta^{\top}D^{\top}W^{-1}D\theta+\sum_{i=1}^n\sum_{j=1}^{N_i}z_{ij}+b_{\sigma}.
\end{align*}
\item The full conditional distribution of $\tau^2$.
\begin{align*}
    p(\tau^2\mid \theta,\sigma^2,w)
    &\propto
    |W|^{-1/2}\exp\left(-\frac{1}{2\sigma^2}\theta^{\top}D^{\top}W^{-1}D\theta\right)(\tau^2)^{-1-1/2}\exp\left(-\frac{1}{\tau^2\xi}\right)\\
    &\propto
    (\tau^2)^{-1-(n-k)/2}\exp\left\{-\frac{1}{\tau^2}\left(\sum_{i=k+2}^n\frac{\eta_i^2}{2\sigma^2w_i^2}+\frac{1}{\xi}\right)\right\},
\end{align*}
which is $\mathrm{IG}\left((n-k)/2, \sum_{i=k+2}^n\eta_i^2/(2\sigma^2w_i^2)+1/\xi\right)$.
\item The full conditional distribution of $\xi$.
\begin{align*}
p(\xi\mid \tau^2)
    &\propto
    \exp\left(-\frac{1}{\tau^2\xi}\right)\xi^{-1-1/2}\exp\left(-\frac{1}{\xi}\right)\\
    &\propto
    \xi^{-1-1/2}\exp\left\{-\frac{1}{\xi}\left(\frac{1}{\tau^2}+1\right)\right\},
\end{align*}
which is $\mathrm{IG}\left(1/2, 1/\tau^2+1\right)$.
\item The full conditional distributions of $w_i^2$ for $i=1,\dots,k+1$.
\begin{align*}
    p(w_i^2\mid \theta_i,\sigma^2)
    &\propto (w_i^2)^{-1/2}\exp\left(-\frac{\eta_i^2}{2w_i^2\sigma^2}\right)(w_i^2)^{-1-a_{w_i}}\exp\left(-\frac{b_{w_i}}{w_i^2}\right)\\
    &\propto (w_i^2)^{-1-(1/2+a_{w_i})}\exp\left(-\left(\frac{\eta_i^2}{2\sigma^2}+b_{w_i}\right)\frac{1}{w_i}\right),
\end{align*}
which is $\mathrm{IG}\left(1/2+a_{w_i}, \eta_i^2/2\sigma^2+b_{w_i}\right)$.
\end{itemize}

The full conditional distributions of $w_i$ for $i=k+2,\dots, n$ and their argumentation parameters are derived separately for Laplace-type and horseshoe-type priors.

\subsection*{Laplace-type}
For $i=k+2,\dots, n$, we assume $w^2_i\sim \mathrm{Exp}(\gamma^2/2)$.
\begin{itemize}
    \item The full conditional distributions of $w_i$ for $i=k+2,\dots,n$.
\begin{align*}
    p(w_i^2\mid \theta_i, \gamma^2)
    &\propto
    |W|^{-1/2}\exp\left(-\frac{1}{2}\sum_{i=1}^{n}\frac{\eta_i^2}{\sigma^2w_i^2}\right)\exp\left(-\frac{\gamma^2}{2}w_i^2\right)\\
    &\propto (w_i^2)^{-1+1/2}\exp\left(-\frac{1}{2}\left(\frac{\eta_i^2}{\sigma^2}\frac{1}{w_i}-\gamma^2w_i^2\right)\right),
\end{align*}
which is $\mathrm{GIG}\left(1/2, \eta_i^2/\sigma^2,\gamma^2\right)$.
\item The full conditional distribution of $\gamma^2$
\begin{align*}
    p(\gamma^2\mid w, \nu^2)
    &\propto
    \prod_{i=k+2}^{n}\left(\frac{\gamma^2}{2}\right)\exp\left(-\frac{\gamma^2}{2}w_i^2\right)(\gamma^2)^{-1-1/2}\exp\left(-\frac{1}{\gamma^2\nu}\right)\\
    &\propto
    (\gamma^2)^{-1+(n-k-1-1/2)}\exp\left(-\frac{1}{2}\left(\gamma^2\sum_{i=k+2}^nw_i^2+\frac{2}{\nu}\frac{1}{\gamma^2}\right)\right),
\end{align*}
which is $\mathrm{GIG}\left(n-k-3/2, 2/\nu,\sum_{i=k+2}^nw_i^2\right)$.
\item The full conditional distribution of $\nu$.
\begin{align*}
    p(\nu\mid \gamma^2)
    &\propto
    \exp\left(-\frac{1}{\gamma^2\nu}\right)\nu^{-1-1/2}\exp\left(-\frac{1}{\nu}\right)\\
    &\propto
    \nu^{-1-1/2}\exp\left(-\left(\frac{1}{\gamma^2}+1\right)\frac{1}{\nu}\right),
\end{align*}
which is $\mathrm{IG}\left(1/2,1/\gamma^2+1\right)$.
\end{itemize}

\subsection*{Horseshoe-type}

For $i=k+2,\dots, n$, we assume $w_i\sim C^{+}(0, 1)$.
\begin{itemize}
\item The full conditional distributions of $w_i$ for $i=k+2,\dots,n$.
\begin{align*}
    p(w_i^2\mid \theta_i)
    &\propto 
    (w_i^2)^{-1/2}\exp\left(-\frac{1}{2\sigma^2}\theta^{\top}D^{\top}W^{-1}D\theta\right)(w_i^2)^{-1-1/2}\exp\left(-\frac{1}{w_i^2\nu_i}\right)\\
    &\propto 
    (w_i^2)^{-1-1}\exp\left(-\left(\frac{\eta_i^2}{2\tau^2\sigma^2}+\frac{1}{\nu_i}\right)\frac{1}{w_i^2}\right),
\end{align*}
which is $\mathrm{IG}\left(1, \eta_i^2/(2\tau^2\sigma^2)+1/\nu_i\right)$.
\item The full conditional distributions of $\nu_i$ for $i=k+2,\dots,n$.
\begin{align*}
    p(\nu_i\mid w_i^2)
    &\propto
    \exp\left(-\frac{1}{w_i^2\nu_i}\right)\nu^{-1-1/2}\exp\left(-\frac{1}{\nu_i}\right)\\
    &\propto
    \nu_i^{-1-1/2}\exp\left(-\left(\frac{1}{w_i^2}+1\right)\frac{1}{\nu_i}\right),
\end{align*}
which is $\mathrm{IG}\left(1/2,1/w_i^2+1\right)$.
\end{itemize}


\section{Variational distributions}

We summarize the derivations of variational distributions.

\begin{itemize}
\item The variational distribution of $\theta$.
\begin{align*}
q(\theta)
&\propto \exp\left( \E_{z,\tau^2,w,\sigma^2}\left[ -\frac{1}{2t^2\sigma^2}\sum_{i=1}^n\sum_{j=1}^{N_i}\left(-\frac{(y_{ij}-\theta_i-\psi z_{ij})^2}{2t^2 \sigma^2z_{ij}}\right)+\left(-\frac{1}{2\sigma^2}\theta^{\top}D^{\top}W^{-1}D\theta\right)\right]\right)\\
&\propto \exp\left(-\frac{1}{2}(\theta-A^{-1}B)^{\top}(\E_{1/\sigma^2}A)(\theta-A^{-1}B)\right),
\end{align*}
which is $N_n(A^{-1}B, (\E_{1/\sigma^2}A)^{-1})$, where
\begin{align*}
    A&=D^{\top}\hat{W}^{-1}D+\frac{1}{t^2}\diag\left(\sum_{j=1}^{N_1}E_{1/z_{1j}},\ldots,\sum_{j=1}^{N_n}E_{1/z_{nj}}\right), \\
    B&=\left(\frac{1}{t^2}\sum_{j=1}^{N_1}\left(y_{1j}E_{1/z_{1j}}-\psi\right),\dots,\frac{1}{t^2}\sum_{j=1}^{N_n}\left(y_{nj}E_{1/z_{nj}}-\psi\right)\right)^{\top}\\
    \hat{W}^{-1}&=\mathrm{diag}(E_{1/w_1^2},\dots,E_{1/w_{k+1}^2},E_{1/\tau^2}E_{1/w_{k+2}^2},\dots,E_{1/\tau^2}E_{1/w_{n}^2}),\\
    E_{1/w_i^2}&=\E_{w_i^2}[1/w_i^2],\quad E_{1/\tau^2}=\E_{\tau^2}[1/\tau^2],\\ E_{1/\sigma^2}&=\E_{\sigma^2}[1/\sigma^2],\quad E_{1/z_{ij}}=\E_{z_{ij}}[1/z_{ij}]
\end{align*}
\item The variational distributions of $z_{ij}$ for $i=1,\dots,n$ and $j=1,\dots,N_j$.
\begin{align*}
    q(z_{ij})
&\propto
\exp\left( \E_{\theta,\tau^2,w,\sigma^2}\left[\log(z_{ij})^{-1/2}+\left(-\frac{(y_{ij}-\theta_i-\psi z_{ij})^2}{2t^2\sigma^2 z_{ij}}\right)+\left(-\frac{z_{ij}}{\sigma^2}\right) \right]\right)\\
&\propto
(z_{ij})^{-1/2}\exp\left(-\frac{1}{2}\left\{ \E_{\sigma^2}\left[\frac{1}{\sigma^2}\right]\E_{\theta}\left[\frac{(y_{ij}-\theta_i)^2}{t^2}\right]\frac{1}{z_{ij}}+\left(\frac{\psi^2}{t^2}-2\right)\E_{\sigma^2}\left[\frac{1}{\sigma^2}\right]z_{ij}\right\}\right),
\end{align*}
which is $\mathrm{GIG}\left(1/2, \alpha_{z_{ij}}, \beta_{z_{ij}}\right)$, where
\begin{align*}
\alpha_{z_{ij}}&=\frac{1}{t^2}E_{1/\sigma^2}(y_{ij}^2-2y_{ij}E_{\theta}+E_{\theta^2}),\quad \beta_{z_{ij}}=\left(\frac{\psi^2}{t^2}-2\right)E_{1/\sigma^2},\\
E_{\theta_i}&=\E_{\theta_i}[\theta_i]=(A^{-1}B)_i,\quad E_{\theta_i^2}=\E_{\theta_i}[\theta_i^2]=e_i^{\top}(E_{\sigma^2}^{-1}A^{-1}+A^{-1}BB^{\top}A^{-1})e_i,
\end{align*}
where $e_i$ is a unit vector that the $i$th component is 1.
\item The variational distribution of $\sigma$.
\begin{align*}
q(\sigma^2)
&\propto
\exp\left( \E_{\theta,\tau^2,w,z}\left[\log(\sigma^2)^{-N/2} + \sum_{i=1}^n\sum_{j=1}^{N_i}\left(-\frac{(y_{ij}-\theta_i-\psi z_{ij})^2}{2t^2\sigma^2 z_{ij}}\right)\right.\right.\\
&\quad\quad \left.\left.+\log(\sigma^2)^{-n/2}-\left(\frac{1}{2\sigma^2}\theta^{\top}D^{\top}W^{-1}D\theta\right)+\log(\sigma^2)^{-N}-\sum_{i=1}^n\sum_{j=1}^N\frac{z_{ij}}{\sigma^2}+\log(\sigma^2)^{-1-a_{\sigma}}-\frac{b_{\sigma}}{\sigma^2} \right]\right)\\
&\propto(\sigma^2)^{-1-(n+3N)/2-a_{\sigma}}\exp\left(-\frac{\alpha_{\sigma^2}}{\sigma^2}\right),
\end{align*}
which is $\mathrm{IG}\left((n+3N)/2+a_{\sigma}, \alpha_{\sigma^2}\right)$, where
\begin{align*}
\alpha_{\sigma^2}
&=\sum_{i=1}^n\sum_{j=1}^{N_i}\frac{1}{2t^2}\left\{\left(y_{ij}^2-2y_{ij}E_{\theta_i}+E_{\theta_i^2}\right)E_{1/z_{ij}}-2\psi\left(y_{ij}-E_{\theta_i}\right)+\psi^2E_{z_{ij}}\right\}\\
&\quad \quad +\frac{1}{2}\left(\sum_{i=1}^{k+1}E_{\eta_i^2}E_{w_i}+\sum_{i=k+2}^nE_{\eta_i^2}E_{1/\tau^2}E_{1/w_i^2}\right)+\sum_{i=1}^n\sum_{j=1}^NE_{z_{ij}}+b_{\sigma},\\
E_{\eta_i^2}&=d_i^{\top}(E_{\sigma^2}^{-1}A^{-1}+A^{-1}BB^{\top}A^{-1})d_i.
\end{align*}
\item The variational distribution of $\tau^2$.
\begin{align*}
    q(\tau^2)
    &\propto \exp\left(\E_{\theta, \sigma^2,w,\xi}\left[\log |W|^{1/2}-\frac{1}{2\sigma^2}\theta^{\top}D^{\top}W^{-1}D\theta +\log (\tau^2)^{-1-1/2}-\frac{1}{\tau^2\xi}\right]\right)\\
    &\propto (\tau^2)^{-1-(n-k)/2}\exp\left\{-\left(\frac{1}{2}E_{1/\sigma^2}\sum_{i=k+2}^nE_{\eta_i^2}E_{1/w_i^2} +E_{1/\xi}\right)\frac{1}{\tau^2}\right\},
\end{align*}
which is $\mathrm{IG}\left((n-k)/2,E_{1/\sigma^2}/2\sum_{i=k+2}^nE_{\eta_i^2}E_{1/w_i^2} +E_{1/\xi} \right)$, where $E_{1/\xi}=\mathrm{E}_{\xi}[1/\xi]$
\item The variational distribution of $\xi$.
\begin{align*}
    q(\xi)
    &\propto\xi^{-1-1/2}\exp\left(-\E_{\tau^2}\left[\frac{1}{\tau^2}\right]\frac{1}{\xi}-\frac{1}{\xi}\right)\\
    &\propto
    \xi^{-1-1/2}\exp\left\{-\left(E_{1/\tau^2}+1\right)\frac{1}{\xi}\right\},
\end{align*}
which is $\mathrm{IG}\left(1/2,E_{1/\tau^2}+1\right)$.
\item The variational distributions of $w_i^2$ for $i=1,\dots,k+1$.
\begin{align*}
    q(w_i^2)
    &\propto \exp\left(\log(w_i^2)^{-1/2}-\E_{\theta, \sigma^2}\left[\frac{\eta_i^2}{2w_i^2\sigma^2}\right]\right)(w_i^2)^{-1-a_{w_i}}\exp\left(-\frac{b_{w_i}}{w_i^2}\right)\\
    &\propto (w_i^2)^{-1-(1/2+a_{w_i})}\exp\left(-\left(\frac{1}{2}E_{\eta_i^2}E_{1/\sigma^2}+b_{w_i}\right)\frac{1}{w_i}\right),
\end{align*}
which is $\mathrm{IG}\left(1/2+a_{w_i}, E_{\eta_i^2}E_{1/\sigma^2}/2+b_{w_i}\right)$.
\end{itemize}

The variational distributions of $w_i$ for $i=k+2,\dots, n$ and their argumentation parameters are derived separately for Laplace-type and horseshoe-type priors.

\subsection*{Laplace-type}
In Laplace-type prior, we set $\tau^2=1$. 
For $i=k+2,\dots, n$, we assume $w^2_i\sim \mathrm{Exp}(\gamma^2/2)$. 
\begin{itemize}
\item The variational distributions of $w_i$ for $i=k+2,\dots,n$.
\begin{align*}
q(w_i^2)
    &\propto
    \exp\left(\log(w_i^2)^{-1/2}-\E_{\theta, \sigma^2}\left[\frac{\eta_i^2}{2w_i^2\sigma^2}\right]-\E_{\gamma^2}\left[\frac{\gamma^2}{2}w_i^2\right]\right)\\
    &\propto (w_i^2)^{-1+1/2}\exp\left(-\frac{1}{2}\left(E_{\eta_i^2}E_{1/\sigma^2}\frac{1}{w_i}+E_{\gamma^2}w_i^2\right)\right),
\end{align*}
which is $\mathrm{GIG}\left(1/2, E_{\eta_i^2}E_{1/\sigma^2},E_{\gamma^2}\right)$.
\item The variational distributions of $\gamma^2$.
\begin{align*}
    q(\gamma^2)
    &\propto
    \exp\left(\log (\gamma^2)^{n-k-1}-\frac{\gamma^2}{2}\sum_{i=k+2}^n\E_{w}[w_i^2]+\log(\gamma^2)^{-1-1/2}-\frac{1}{\gamma^2}\E_{\nu}\left[\frac{1}{\nu}\right]\right)\\
    &\propto
    (\gamma^2)^{-1+(n-k-1-1/2)}\exp\left(-\frac{1}{2}\left(\gamma^2\sum_{i=k+2}^nE_{w_i^2}+2E_{1/\nu}\frac{1}{\gamma^2}\right)\right),
\end{align*}
which is $\mathrm{GIG}\left(n-k-3/2, 2E_{1/\nu},\sum_{i=k+2}^nE_{w_i^2}\right)$, where $E_{1/\nu}=\mathrm{E}_{\nu}[1/\nu]$.
\item The variational distributions of $\nu$.
\begin{align*}
    q(\nu)
    &\propto
    \exp\left(-\E_{\gamma^2}\left[\frac{1}{\gamma^2}\right]\frac{1}{\nu}\right)\nu^{-1-1/2}\exp\left(-\frac{1}{\nu}\right)\\
    &\propto
    \nu^{-1-1/2}\exp\left(-\left(E_{1/\gamma^2}+1\right)\frac{1}{\nu}\right),
\end{align*}
which is $\mathrm{IG}\left(1/2,E_{1/\gamma^2}+1\right)$, where $E_{1/\gamma^2}=\mathrm{E}_{\gamma^2}[1/\gamma^2]$.
\end{itemize}

\subsection*{Horseshoe-type}

For $i=k+2,\dots, n$, we assume $w_i\sim C^{+}(0, 1)$. 
\begin{itemize}
\item The variational distributions of $w_i$ ($i=k+2,\dots,n$).
\begin{align*}
    q(w_i^2)
    &\propto 
    \exp\left(\log (w_i^2)^{-1/2}-\frac{1}{2}\E_{\theta,\tau^2,\sigma^2}\left[\frac{1}{\sigma^2}\theta^{\top}D^{\top}W^{-1}D\theta\right]+\log(w_i^2)^{-1-1/2}-\E_{\nu_i}\left[\frac{1}{\nu_i}\right]\frac{1}{w_i^2}\right)\\
    &\propto 
    (w_i^2)^{-1-1}\exp\left(-\left(\frac{1}{2}E_{\eta_i^2}E_{1/\tau^2}E_{1/\sigma^2}+E_{1/\nu_i}\right)\frac{1}{w_i^2}\right),
\end{align*}
which is $\mathrm{IG}\left(1, E_{\eta_i^2}E_{1/\tau^2}E_{1/\sigma^2}/2+E_{1/\nu_i}\right)$, where $E_{1/\nu_i}=\mathrm{E}_{\nu_i}[1/\nu_i]$.
\item The variational distributions of $\nu_i$ ($i=k+2,\dots,n$).
\begin{align*}
    q(\nu_i)
    &\propto
    \exp\left(-\E_{w_i^2}\left[\frac{1}{w_i^2}\right]\frac{1}{\nu_i}\right)\nu_i^{-1-1/2}\exp\left(-\frac{1}{\nu_i}\right)\\
    &\propto
    \nu_i^{-1-1/2}\exp\left(-\left(E_{1/w_i^2}+1\right)\frac{1}{\nu_i}\right),
\end{align*}
which is $\mathrm{IG}\left(1/2,E_{1/w_i^2}+1\right)$.
\end{itemize}

\section{Additional information on simulation studies}

\subsection{Smooth Gaussian process}\label{supp:sec:3.1}

\begin{figure}[thbp]
\begin{center}
\includegraphics[width=\linewidth]{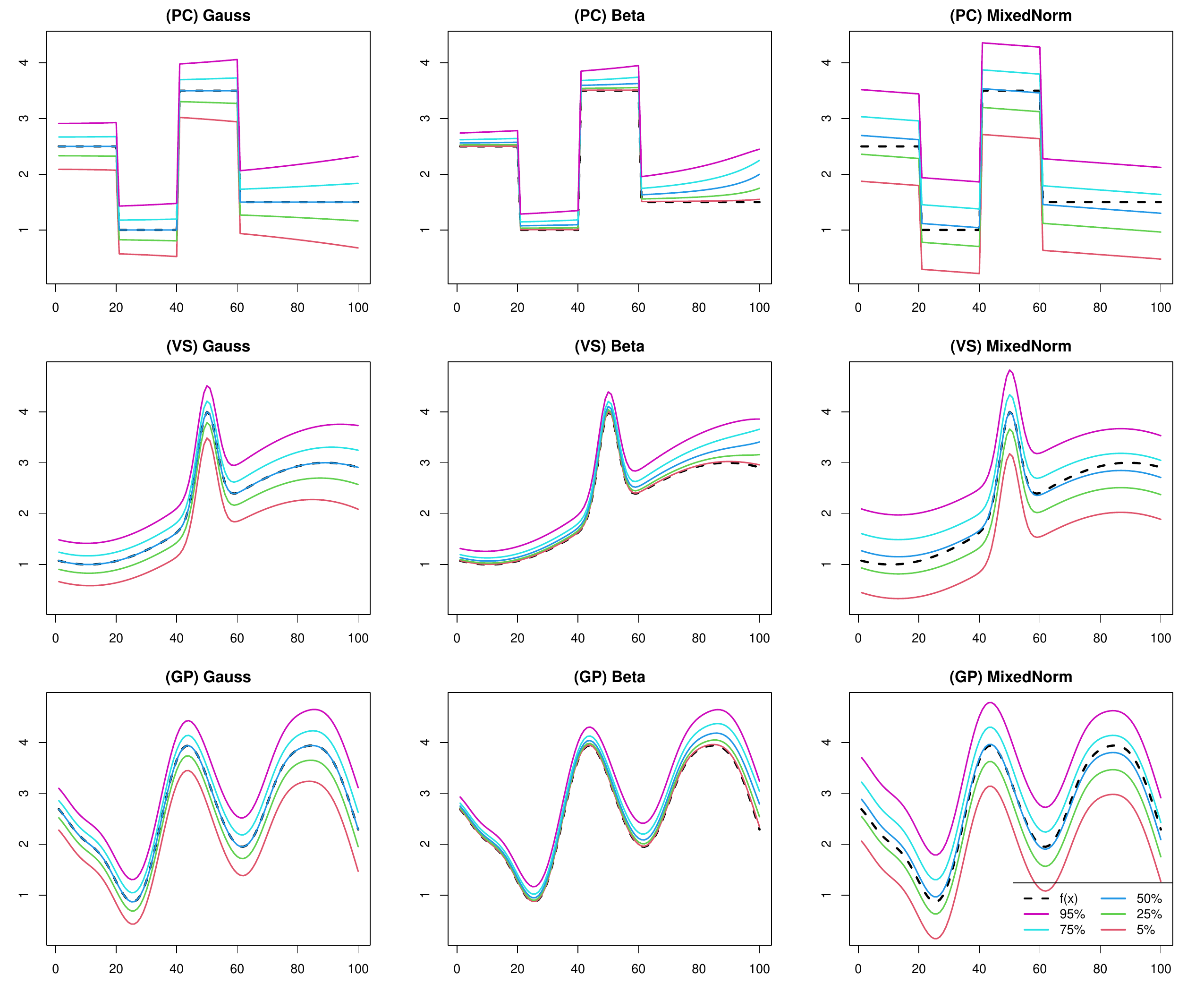}
\caption{Simulated true quantile trends.}
\label{true}
\end{center}
\end{figure}

In Section 4 of the main manuscript, we provided simulation results under (PC) piecewise constant and (VS) varying smoothness. In this subsection, we provide simulation results under the smooth Gaussian process (denoted by (GP)) as a true data-generating scenario. The plots of simulated true quantiles are shown in Figure \ref{true}. The additional description and results under (GP) are as follows.
\begin{itemize}
\item[(GP)] Smooth Gaussian process
\begin{align*}
f\sim \mathrm{GP}(\mu,\Sigma),\quad \Sigma_{i,j}=\sigma_f^2\exp\{-(t_j-t_i)^2/(2\rho^2)\}.
\end{align*}
\end{itemize}
The scenario (GP) generates observations from the Gaussian process with squared exponential covariance function \citep[see also ][]{faulkner2018locally}. We set $\mu=2$, $\sigma_f^2=1$ and $\rho=10$. The function $f$ was generated with the same random number seed for all scenarios (Gaussian, Beta, and Mixed normal noise functions). The aim of the scenario (GP) is to test the ability of the proposed methods to handle a smoothly varying function with no local change. To this end, we consider $k=2$ in this scenario.  

The results are represented in Table~\ref{GP_table_MSEMADMCIWCP}. The raw computing time and effective sample size per unit time are also reported in Table~\ref{GP_table_timeESS}. Note that the simulation is conducted without parallel computation as well as those of Section 4 in the main text.

\begin{table}[thbp]
\caption{Average values of MSE, MAD, MCIW and CP based on $100$ replications for Gaussian process with $k=2$. The minimum values and second smallest values of MSE and MAD are represented in bold and italics respectively. The CP values above 90\% are represented in bold. }
\begin{center}
\resizebox{1.0\textwidth}{!}{ 
\begin{tabular}{c|ccccc|ccccc}
\toprule
& \multicolumn{5}{c|}{MSE} &  \multicolumn{5}{c}{MAD } \\
\hline
(I) Gauss & 0.05 & 0.25 & 0.5 & 0.75 & 0.95 & 0.05 & 0.25 & 0.5 & 0.75 & 0.95 \\ 
  \hline
MCMC-HS & 0.084 & {\sl 0.020} & {\bf 0.016} & {\bf 0.019} & {\bf 0.040} & {\sl 0.187} & {\sl 0.109} & {\bf 0.099} & {\bf 0.107} & {\bf 0.153} \\ 
  VB-HS & 0.174 & 0.024 & 0.019 & 0.023 & 0.048 & 0.243 & 0.118 & 0.107 & 0.117 & 0.168 \\ 
  MCMC-Lap & 0.093 & 0.021 & {\sl 0.017} & {\sl 0.020} & {\sl 0.043} & 0.195 & {\sl 0.110} & 0.100 & {\sl 0.108} & {\sl 0.159} \\ 
  VB-Lap & {\bf 0.048} & {\bf 0.019} & {\sl 0.017} & {\sl 0.020} & 0.048 & {\bf 0.170} & {\bf 0.107} & {\sl 0.100} & 0.110 & 0.171 \\ 
  ADMM & {\sl 0.050} & 0.024 & 0.023 & 0.026 & 0.049 & 0.174 & 0.120 & 0.115 & 0.123 & 0.172 \\ 
\midrule
\midrule
(II) Beta & 0.05 & 0.25 & 0.5 & 0.75 &0.95 & 0.05 & 0.25 & 0.5 & 0.75 &0.95  \\
\hline
MCMC-HS & {\sl 0.004} & {\bf 0.003} & {\bf 0.005} & {\bf 0.007} & {\bf 0.014} & 0.041 & {\sl 0.037} & {\bf 0.048} & {\sl 0.062} & {\bf 0.092} \\ 
  VB-HS & 0.009 & {\sl 0.004} & {\sl 0.006} & {\sl 0.008} & 0.017 & {\sl 0.037} & 0.041 & 0.054 & 0.067 & 0.101 \\ 
  MCMC-Lap & {\bf 0.003} & {\bf 0.003} & {\bf 0.005} & {\bf 0.007} &  {\sl 0.015} & 0.039 & {\bf 0.036} & {\bf 0.048} & {\bf 0.061} & {\sl 0.096} \\ 
  VB-Lap & {\bf 0.003} & {\bf 0.003} & {\bf 0.005} & {\bf 0.007} & 0.018 & {\bf 0.030} & {\bf 0.036} & {\sl 0.049} & 0.063 & 0.104 \\ 
  ADMM & {\sl 0.004} & {\sl 0.004} & {\sl 0.006} & 0.009 & 0.017 & 0.040 & 0.039 & 0.055 & 0.069 & 0.103 \\ 
 \midrule
\midrule
(III) {\small Mixed normal} & 0.05 & 0.25 & 0.5 & 0.75 &0.95 & 0.05 & 0.25 & 0.5 & 0.75 &0.95  \\
\hline
MCMC-HS & 0.169 & 0.064 & {\sl 0.037} & {\sl 0.041} & {\bf 0.089} & 0.278 & {\sl 0.177} & {\sl 0.150} & {\bf 0.158} & {\bf 0.234} \\ 
  VB-HS & 0.205 & 0.087 & 0.043 & 0.047 & 0.106 & 0.309 & 0.196 & 0.159 & 0.169 & 0.254 \\ 
  MCMC-Lap & 0.171 & 0.067 & {\sl 0.037} & {\bf 0.040} & {\sl 0.094} & 0.282 & 0.178 & {\bf 0.149} & {\bf 0.158} & {\sl 0.242} \\ 
  VB-Lap & {\bf 0.100} & {\bf 0.043} & {\bf 0.035} & {\sl 0.041} & 0.103 & {\bf 0.250} & {\bf 0.164} & {\bf 0.149} & {\sl 0.160} & 0.252 \\ 
  ADMM & {\sl 0.106} & {\sl 0.056} & 0.046 & 0.053 & 0.111 & {\sl 0.259} & 0.187 & 0.170 & 0.183 & 0.261 \\
\midrule
\multicolumn{11}{c}{}\\
\midrule
& \multicolumn{5}{c|}{MCIW} &  \multicolumn{5}{c}{CP} \\
\hline
(I) Gauss & 0.05 & 0.25 & 0.5 & 0.75 & 0.95 & 0.05 & 0.25 & 0.5 & 0.75 & 0.95 \\ 
  \hline
MCMC-HS & 0.510 & 0.478 & 0.462 & 0.458 & 0.414 & 0.738 & {\bf 0.906} & {\bf 0.935} & {\bf 0.908} & 0.723 \\ 
  CVB-HS & 1.267 & 0.616 & 0.486 & 0.599 & 1.161 & {\bf 0.950} & {\bf 0.943} & {\bf 0.921} & {\bf 0.941} & {\bf 0.980} \\ 
  VB-HS & 0.145 & 0.217 & 0.232 & 0.216 & 0.124 & 0.221 & 0.546 & 0.610 & 0.538 & 0.235 \\ 
  MCMC-Lap & 0.520 & 0.494 & 0.478 & 0.470 & 0.422 & 0.748 & {\bf 0.913} & {\bf 0.944} & {\bf 0.916} & 0.720 \\ 
  CVB-Lap & 1.534 & 0.864 & 0.589 & 0.792 & 1.226 & {\bf 0.994} & {\bf 0.996} & {\bf 0.972} & {\bf 0.991} & {\bf 0.982} \\ 
  VB-Lap & 0.185 & 0.291 & 0.309 & 0.289 & 0.182 & 0.333 & 0.721 & 0.777 & 0.713 & 0.330 \\
\midrule
\midrule
(II) Beta & 0.05 & 0.25 & 0.5 & 0.75 &0.95 & 0.05 & 0.25 & 0.5 & 0.75 &0.95  \\
\hline
MCMC-HS & 0.161 & 0.179 & 0.214 & 0.244 & 0.233 & {\bf 0.948} & {\bf 0.946} & {\bf 0.923} & 0.891 & 0.693 \\ 
  CVB-HS & 0.490 & 0.272 & 0.214 & 0.350 & 0.725 & {\bf 0.991} & {\bf 0.962} & 0.885 & {\bf 0.933} & {\bf 0.972} \\ 
  VB-HS & 0.045 & 0.087 & 0.109 & 0.111 & 0.066 & 0.474 & 0.674 & 0.621 & 0.507 & 0.215 \\ 
  MCMC-Lap & 0.161 & 0.182 & 0.221 & 0.252 & 0.239 & {\bf 0.956} & {\bf 0.950} & {\bf 0.936} & {\bf 0.902} & 0.692 \\ 
  CVB-Lap & 0.608 & 0.377 & 0.271 & 0.459 & 0.737 & {\bf 0.993} & {\bf 0.990} & {\bf 0.951} & {\bf 0.981} & {\bf 0.977} \\ 
  VB-Lap & 0.069 & 0.119 & 0.149 & 0.154 & 0.102 & 0.770 & 0.855 & 0.808 & 0.697 & 0.307 \\ 
 \midrule
\midrule
(III) {\small Mixed normal} & 0.05 & 0.25 & 0.5 & 0.75 &0.95 & 0.05 & 0.25 & 0.5 & 0.75 &0.95  \\
\hline
MCMC-HS & 0.674 & 0.740 & 0.706 & 0.680 & 0.628 & 0.693 & {\bf 0.905} & {\bf 0.937} & {\bf 0.901} & 0.728 \\ 
  CVB-HS & 1.652 & 0.921 & 0.746 & 0.847 & 1.650 & {\bf 0.966} & {\bf 0.939} & {\bf 0.929} & {\bf 0.934} & {\bf 0.976} \\ 
  VB-HS & 0.203 & 0.329 & 0.348 & 0.317 & 0.193 & 0.218 & 0.515 & 0.609 & 0.559 & 0.236 \\ 
  MCMC-Lap & 0.693 & 0.758 & 0.725 & 0.702 & 0.646 & 0.710 & {\bf 0.913} & {\bf 0.946} & {\bf 0.916} & 0.724 \\ 
  CVB-Lap & 1.909 & 1.145 & 0.878 & 1.134 & 1.783 & {\bf 0.993} & {\bf 0.995} & {\bf 0.981} & {\bf 0.993} & {\bf 0.985} \\ 
  VB-Lap & 0.274 & 0.425 & 0.452 & 0.421 & 0.275 & 0.324 & 0.693 & 0.769 & 0.697 & 0.333 \\
\bottomrule
\end{tabular}}
\end{center}
\label{GP_table_MSEMADMCIWCP}
\end{table}

\begin{table}[thbp]
\caption{Average values of raw computing time and effective sample size per unit time based on $100$ replications for GP scenarios.}
\begin{center}
\resizebox{1.0\textwidth}{!}{ 
\begin{tabular}{c|ccccc|ccccc}
\toprule
& \multicolumn{5}{c|}{Computation time (second)} &  \multicolumn{5}{c}{ESS (per second)} \\
\midrule
(I) Gauss & 0.05 & 0.25 & 0.5 & 0.75 & 0.95 & 0.05 & 0.25 & 0.5 & 0.75 & 0.95 \\ 
  \hline
MCMC-HS & 32 & 32 & 32 & 32 & 32 & 14 & 38 & 44 & 41 & 15 \\ 
  CVB-HS & 13 & 8 & 11 & 8 & 13 & 575 & 943 & 692 & 980 & 576 \\ 
  MCMC-Lap & 35 & 35 & 35 & 35 & 35 & 13 & 56 & 69 & 59 & 14 \\ 
  CVB-Lap & 20 & 17 & 17 & 12 & 16 & 378 & 442 & 431 & 627 & 475 \\
\midrule
\midrule
(II) Beta & 0.05 & 0.25 & 0.5 & 0.75 &0.95 & 0.05 & 0.25 & 0.5 & 0.75 &0.95  \\
\hline
MCMC-HS & 32 & 32 & 32 & 32 & 32 & 13 & 48 & 45 & 42 & 14 \\ 
  CVB-HS & 9 & 7 & 8 & 7 & 11 & 859 & 1144 & 907 & 1113 & 714 \\ 
  MCMC-Lap & 35 & 35 & 35 & 35 & 35 & 11 & 73 & 88 & 59 & 13 \\ 
  CVB-Lap & 30 & 8 & 12 & 8 & 10 & 257 & 915 & 639 & 997 & 728 \\  
 \midrule
\midrule
(III) {\small Mixed normal} & 0.05 & 0.25 & 0.5 & 0.75 &0.95 & 0.05 & 0.25 & 0.5 & 0.75 &0.95  \\
\hline
MCMC-HS & 32 & 32 & 32 & 32 & 32 & 15 & 36 & 42 & 39 & 15 \\ 
  CVB-HS & 15 & 10 & 14 & 9 & 16 & 503 & 732 & 526 & 849 & 487 \\ 
  MCMC-Lap & 35 & 35 & 35 & 35 & 35 & 14 & 51 & 59 & 53 & 14 \\ 
  CVB-Lap & 29 & 32 & 25 & 20 & 24 & 259 & 241 & 304 & 375 & 320 \\ 
\bottomrule
\end{tabular}}
\end{center}
\label{GP_table_timeESS}
\end{table}

\subsection{Multiple observations}

We provide simulation results under multiple observations per location. The setting of simulation is as follows. Data was simulated by the equation (5) in the main manuscript, where the number of data at each location $N_i$ ($i=1,\dots,n$) was simulated from a multinomial distribution with outcomes 2, 3, and 4 with probabilities $1/3$. We used the same noise distributions as those of Section 4 in the main manuscript (Gauss, Beta, and Mixed normal). The number of locations was set as $n=100$, while the total sample size $N$ was 200 to 400 because of the randomness of $N_i$. As true data-generating functions, (PC), (VS) and (GP) were adopted. The convergence criterion of variational Bayes methods was set as $10^{-3}$, which is slightly different from that of the main simulation. We compared the proposed methods with the quantile smoothing spline method (denoted by Spline for short) by \cite{nychka2017fields} as a frequentist method, which was also applied to Munich rent data in Section 5.2 of the main manuscript. The reason is that Brantley's quantile trend filtering cannnot apply to multiple observations per location. 

The results of the simulations are summarized in Tables \ref{MPC_table_MSEMADMCIWCP}, \ref{MVS_table_MSEMADMCIWCP} and \ref{MGP_table_MSEMADMCIWCP}. From these results, the MCMC-HS method was the best and the VB-HS method was at least better than the Spline method for almost all scenarios in terms of point estimation. 
Compared with the results of a single observation in the main manuscript, the MSE and MAD were smaller for each method, and it seems that a larger sample size induces more accurate point estimation.
Although the CVB method tended to be over-coverage, it gave wider credible intervals than those of the MCMC method as seen in the MCIW, and conservatively improved coverage probability as well as single observation.

\begin{table}[thbp]
\caption{Average values of MSE, MAD, MCIW and CP based on $100$ replications for multiple observations (PC) with $k=0$. The minimum values and second smallest values of MSE and MAD are represented in bold and italics respectively. The CP values above 90\% are represented in bold. }
\begin{center}
\resizebox{1.0\textwidth}{!}{ 
\begin{tabular}{c|ccccc|ccccc}
\toprule
& \multicolumn{5}{c|}{MSE} &  \multicolumn{5}{c}{MAD } \\
\hline
(I) Gauss & 0.05 & 0.25 & 0.5 & 0.75 & 0.95 & 0.05 & 0.25 & 0.5 & 0.75 & 0.95 \\ 
  \hline
MCMC-HS & {\bf 0.0322} & {\bf 0.0052} & {\bf 0.0033} & {\bf 0.0049} & {\bf 0.0332} & {\bf 0.1350} & {\bf 0.0535} & {\bf 0.0436} & {\bf 0.0518} & {\bf 0.1387} \\ 
  VB-HS & {\sl 0.0387} & {\sl 0.0182} & {\sl 0.0132} & {\sl 0.0180} & {\sl 0.0400} & {\sl 0.1516} & {\sl 0.0971} & {\sl 0.0812} & {\sl 0.0946} & {\sl 0.1564} \\ 
  MCMC-Lap & 0.0582 & 0.0205 & 0.0167 & 0.0208 & 0.0576 & 0.1917 & 0.1110 & 0.0996 & 0.1111 & 0.1921 \\ 
  VB-Lap & 0.0589 & 0.0190 & 0.0154 & 0.0190 & 0.0587 & 0.1950 & 0.1060 & 0.0950 & 0.1053 & 0.1974 \\ 
  Spline & 0.1154 & 0.0671 & 0.0548 & 0.0672 & 0.1141 & 0.2214 & 0.1586 & 0.1473 & 0.1582 & 0.2247 \\ 
\midrule
\midrule
(II) Beta & 0.05 & 0.25 & 0.5 & 0.75 &0.95 & 0.05 & 0.25 & 0.5 & 0.75 &0.95  \\
\hline
MCMC-HS & {\sl 0.0011} & {\bf 0.0013} & {\bf 0.0022} & {\bf 0.0037} & {\bf 0.0155} & 0.0180 & {\bf 0.0177} & {\bf 0.0276} & {\bf 0.0412} & {\bf 0.0948} \\ 
  VB-HS & {\bf 0.0008} & {\sl 0.0029} & {\sl 0.0048} & {\sl 0.0079} & {\sl 0.0162} & {\bf 0.0109} & {\sl 0.0270} & {\sl 0.0416} & {\sl 0.0606} & {\sl 0.0991} \\ 
  MCMC-Lap & 0.0046 & 0.0040 & 0.0056 & 0.0092 & 0.0247 & 0.0433 & 0.0398 & 0.0514 & 0.0702 & 0.1236 \\ 
  VB-Lap & 0.0042 & 0.0040 & 0.0056 & 0.0093 & 0.0308 & 0.0351 & 0.0386 & 0.0511 & 0.0709 & 0.1420 \\ 
  Spline & 0.1120 & 0.0580 & 0.0424 & 0.0659 & 0.1042 & 0.1240 & 0.0992 & 0.1021 & 0.1257 & 0.1736 \\ 
 \midrule
\midrule
(III) {\small Mixed normal} & 0.05 & 0.25 & 0.5 & 0.75 &0.95 & 0.05 & 0.25 & 0.5 & 0.75 &0.95  \\
\hline
MCMC-HS & {\bf 0.0656} & {\bf 0.0136} & {\bf 0.0094} & {\bf 0.0143} & {\bf 0.0669} & {\bf 0.2004} & {\bf 0.0889} & {\bf 0.0728} & {\bf 0.0885} & {\bf 0.2016} \\ 
  VB-HS & {\sl 0.0822} & {\sl 0.0397} & {\sl 0.0281} & {\sl 0.0406} & {\sl 0.0827} & {\sl 0.2277} & {\sl 0.1507} & {\sl 0.1239} & {\sl 0.1519} & {\sl 0.2286} \\ 
  MCMC-Lap & 0.1124 & 0.0433 & 0.0355 & 0.0443 & 0.1129 & 0.2754 & 0.1642 & 0.1473 & 0.1650 & 0.2745 \\ 
  VB-Lap & 0.1043 & 0.0381 & 0.0313 & 0.0397 & 0.1061 & 0.2649 & 0.1526 & 0.1372 & 0.1545 & 0.2658 \\ 
  Spline & 0.1559 & 0.0866 & 0.0763 & 0.0871 & 0.1595 & 0.2915 & 0.2053 & 0.1919 & 0.2067 & 0.2958 \\ 
\midrule
\multicolumn{11}{c}{}\\
\midrule
& \multicolumn{5}{c|}{MCIW} &  \multicolumn{5}{c}{CP} \\
\hline
(I) Gauss & 0.05 & 0.25 & 0.5 & 0.75 & 0.95 & 0.05 & 0.25 & 0.5 & 0.75 & 0.95 \\ 
  \hline
MCMC-HS & 0.391 & 0.271 & 0.246 & 0.272 & 0.394 & 0.723 & {\bf 0.938} & {\bf 0.960} & {\bf 0.947} & 0.720 \\ 
  CVB-HS & 1.097 & 0.664 & 0.429 & 0.675 & 1.125 & {\bf 0.983} & {\bf 0.991} & {\bf 0.967} & {\bf 0.994} & {\bf 0.985} \\ 
  VB-HS & 0.108 & 0.181 & 0.187 & 0.184 & 0.110 & 0.228 & 0.568 & 0.675 & 0.602 & 0.219 \\ 
  MCMC-Lap & 0.476 & 0.533 & 0.532 & 0.537 & 0.476 & 0.734 & {\bf 0.941} & {\bf 0.964} & {\bf 0.943} & 0.732 \\ 
  CVB-Lap & 1.296 & 0.678 & 0.497 & 0.690 & 1.323 & {\bf 0.981} & {\bf 0.979} & {\bf 0.953} & {\bf 0.982} & {\bf 0.985} \\ 
  VB-Lap & 0.203 & 0.342 & 0.362 & 0.344 & 0.204 & 0.288 & 0.804 & 0.871 & 0.809 & 0.275 \\
\midrule
\midrule
(II) Beta & 0.05 & 0.25 & 0.5 & 0.75 &0.95 & 0.05 & 0.25 & 0.5 & 0.75 &0.95  \\
\hline
MCMC-HS & 0.054 & 0.084 & 0.124 & 0.171 & 0.216 & {\bf 0.972} & {\bf 0.951} & {\bf 0.932} & {\bf 0.902} & 0.606 \\ 
  CVB-HS & 0.454 & 0.266 & 0.194 & 0.425 & 0.748 & {\bf 1.000} & {\bf 0.987} & {\bf 0.934} & {\bf 0.985} & {\bf 0.983} \\ 
  VB-HS & 0.026 & 0.065 & 0.086 & 0.097 & 0.056 & 0.845 & 0.795 & 0.702 & 0.543 & 0.184 \\ 
  MCMC-Lap & 0.171 & 0.205 & 0.247 & 0.276 & 0.238 & {\bf 0.942} & {\bf 0.950} & {\bf 0.939} & 0.889 & 0.619 \\ 
  CVB-Lap & 0.496 & 0.311 & 0.231 & 0.451 & 0.797 & {\bf 0.992} & {\bf 0.976} & {\bf 0.912} & {\bf 0.970} & {\bf 0.964} \\ 
  VB-Lap & 0.070 & 0.136 & 0.172 & 0.177 & 0.090 & 0.737 & 0.863 & 0.836 & 0.710 & 0.161 \\ 
 \midrule
\midrule
(III) {\small Mixed normal} & 0.05 & 0.25 & 0.5 & 0.75 &0.95 & 0.05 & 0.25 & 0.5 & 0.75 &0.95  \\
\hline
MCMC-HS & 0.628 & 0.439 & 0.404 & 0.439 & 0.615 & 0.750 & {\bf 0.935} & {\bf 0.957} & {\bf 0.927} & 0.742 \\ 
  CVB-HS & 1.520 & 1.007 & 0.677 & 1.009 & 1.551 & {\bf 0.985} & {\bf 0.996} & {\bf 0.978} & {\bf 0.993} & {\bf 0.985} \\ 
  VB-HS & 0.169 & 0.272 & 0.276 & 0.273 & 0.169 & 0.229 & 0.524 & 0.632 & 0.527 & 0.217 \\ 
  MCMC-Lap & 0.734 & 0.792 & 0.778 & 0.790 & 0.727 & 0.746 & 0.944 & 0.962 & 0.941 & 0.741 \\ 
  CVB-Lap & 1.789 & 0.950 & 0.702 & 0.958 & 1.809 & {\bf 0.987} & {\bf 0.982} & {\bf 0.953} & {\bf 0.979} & {\bf 0.987} \\ 
  VB-Lap & 0.319 & 0.497 & 0.517 & 0.496 & 0.318 & 0.336 & 0.806 & 0.867 & 0.805 & 0.339 \\ 
\bottomrule
\end{tabular}}
\end{center}
\label{MPC_table_MSEMADMCIWCP}
\end{table}

\begin{table}[thbp]
\caption{Average values of MSE, MAD, MCIW and CP based on $100$ replications for multiple observations (VS) with $k=1$. The minimum values and second smallest values of MSE and MAD are represented in bold and italics respectively. The CP values above 90\% are represented in bold. }
\begin{center}
\resizebox{1.0\textwidth}{!}{ 
\begin{tabular}{c|ccccc|ccccc}
\toprule
& \multicolumn{5}{c|}{MSE} &  \multicolumn{5}{c}{MAD } \\
\hline
(I) Gauss & 0.05 & 0.25 & 0.5 & 0.75 & 0.95 & 0.05 & 0.25 & 0.5 & 0.75 & 0.95 \\ 
  \hline
MCMC-HS & {\bf 0.0246} & {\bf 0.0079} & {\bf 0.0063} & {\bf 0.0073} & {\bf 0.0255} & {\bf 0.1168} & {\bf 0.0667} & {\bf 0.0597} & {\bf 0.0638} & {\bf 0.1192} \\ 
  VB-HS & {\sl 0.0313} & 0.0135 & 0.0104 & 0.0138 & {\sl 0.0331} & {\sl 0.1325} & 0.0846 & {\sl 0.0749} & 0.0846 & {\sl 0.1370} \\ 
  MCMC-Lap & 0.0346 & 0.0117 & 0.0099 & 0.0113 & 0.0351 & 0.1438 & 0.0838 & 0.0777 & 0.0828 & 0.1458 \\ 
  VB-Lap & 0.0320 & {\sl 0.0115} & {\sl 0.0098} & {\sl 0.0112} & 0.0334 & 0.1373 & {\sl 0.0831} & 0.0773 & {\sl 0.0821} & 0.1415 \\ 
  Spline & 0.0400 & 0.0151 & 0.0126 & 0.0149 & 0.0423 & 0.1558 & 0.0955 & 0.0878 & 0.0949 & 0.1615 \\ 
\midrule
\midrule
(II) Beta & 0.05 & 0.25 & 0.5 & 0.75 &0.95 & 0.05 & 0.25 & 0.5 & 0.75 &0.95  \\
\hline
MCMC-HS & 0.0016 & {\bf 0.0017} & {\bf 0.0024} & {\bf 0.0036} & {\bf 0.0118} & 0.0254 & {\bf 0.0249} & {\bf 0.0329} & {\bf 0.0426} & {\bf 0.0817} \\ 
  VB-HS & 0.0017 & 0.0026 & 0.0042 & 0.0065 & {\sl 0.0122} & {\sl 0.0209} & 0.0289 & 0.0414 & 0.0561 & {\sl 0.0842} \\ 
  MCMC-Lap & {\sl 0.0015} & {\sl 0.0020} & {\sl 0.0032} & {\sl 0.0051} & 0.0159 & 0.0269 & 0.0284 & 0.0389 & 0.0526 & 0.0977 \\ 
  VB-Lap & {\bf 0.0013} & {\sl 0.0020} & {\sl 0.0032} & {\sl 0.0051} & 0.0142 & {\bf 0.0199} & {\sl 0.0282} & {\sl 0.0392} & {\sl 0.0527} & 0.0930 \\ 
  Spline & 0.0016 & 0.0022 & 0.0036 & 0.0057 & 0.0144 & 0.0239 & 0.0310 & 0.0430 & 0.0567 & 0.0949 \\
 \midrule
\midrule
(III) {\small Mixed normal} & 0.05 & 0.25 & 0.5 & 0.75 &0.95 & 0.05 & 0.25 & 0.5 & 0.75 &0.95  \\
\hline
MCMC-HS & {\bf 0.0494} & {\bf 0.0191} & {\bf 0.0152} & {\bf 0.0178} & {\bf 0.0514} & {\bf 0.1702} & {\bf 0.1048} & {\bf 0.0914} & {\bf 0.1007} & {\bf 0.1758} \\ 
  VB-HS & {\sl 0.0649} & 0.0282 & {\sl 0.0226} & 0.0275 & 0.0677 & {\sl 0.1979} & {\sl 0.1274} & {\sl 0.1119} & {\sl 0.1253} & {\sl 0.2030} \\ 
  MCMC-Lap & 0.0695 & {\sl 0.0280} & 0.0229 & {\sl 0.0255} & 0.0732 & 0.2092 & 0.1298 & 0.1174 & 0.1265 & 0.2153 \\ 
  VB-Lap & 0.0651 & 0.0287 & 0.0228 & 0.0257 & {\sl 0.0664} & 0.2003 & 0.1304 & 0.1164 & 0.1259 & 0.2034 \\ 
  Spline & 0.0925 & 0.0354 & 0.0291 & 0.0351 & 0.0917 & 0.2422 & 0.1501 & 0.1352 & 0.1493 & 0.2422 \\ 
\midrule
\multicolumn{11}{c}{}\\
\midrule
& \multicolumn{5}{c|}{MCIW} &  \multicolumn{5}{c}{CP} \\
\hline
(I) Gauss & 0.05 & 0.25 & 0.5 & 0.75 & 0.95 & 0.05 & 0.25 & 0.5 & 0.75 & 0.95 \\ 
  \hline
MCMC-HS & 0.337 & 0.277 & 0.275 & 0.278 & 0.341 & 0.731 & 0.897 & {\bf 0.921} & {\bf 0.913} & 0.734 \\ 
  CVB-HS & 1.134 & 0.592 & 0.403 & 0.617 & 1.172 & {\bf 0.994} & {\bf 0.989} & {\bf 0.965} & {\bf 0.994} & {\bf 0.994} \\ 
  VB-HS & 0.104 & 0.154 & 0.163 & 0.159 & 0.105 & 0.260 & 0.556 & 0.645 & 0.584 & 0.254 \\ 
  MCMC-Lap & 0.403 & 0.373 & 0.370 & 0.374 & 0.400 & 0.740 & {\bf 0.918} & {\bf 0.944} & {\bf 0.925} & 0.734 \\ 
  CVB-Lap & 1.150 & 0.580 & 0.425 & 0.599 & 1.184 & {\bf 0.993} & {\bf 0.989} & {\bf 0.967} & {\bf 0.993} & {\bf 0.993} \\ 
  VB-Lap & 0.170 & 0.228 & 0.240 & 0.230 & 0.172 & 0.387 & 0.735 & 0.787 & 0.743 & 0.378 \\ 
\midrule
\midrule
(II) Beta & 0.05 & 0.25 & 0.5 & 0.75 &0.95 & 0.05 & 0.25 & 0.5 & 0.75 &0.95  \\
\hline
MCMC-HS & 0.084 & 0.110 & 0.136 & 0.162 & 0.200 & {\bf 0.940} & {\bf 0.931} & {\bf 0.904} & 0.864 & 0.648 \\ 
  CVB-HS & 0.082 & 0.131 & 0.191 & 0.233 & 0.237 & 0.887 & {\bf 0.927} & {\bf 0.931} & {\bf 0.900} & 0.738 \\ 
  VB-HS & 0.030 & 0.067 & 0.085 & 0.090 & 0.057 & 0.655 & 0.747 & 0.670 & 0.538 & 0.226 \\ 
  MCMC-Lap & 0.117 & 0.146 & 0.182 & 0.213 & 0.225 & {\bf 0.950} & {\bf 0.951} & {\bf 0.934} & 0.891 & 0.648 \\ 
  CVB-Lap & 0.514 & 0.280 & 0.213 & 0.394 & 0.781 & {\bf 0.999} & {\bf 0.990} & {\bf 0.951} & {\bf 0.989} & {\bf 0.993} \\ 
  VB-Lap & 0.052 & 0.097 & 0.121 & 0.130 & 0.093 & 0.790 & 0.857 & 0.808 & 0.707 & 0.312 \\ 
 \midrule
\midrule
(III) {\small Mixed normal} & 0.05 & 0.25 & 0.5 & 0.75 &0.95 & 0.05 & 0.25 & 0.5 & 0.75 &0.95  \\
\hline
MCMC-HS & 0.486 & 0.418 & 0.406 & 0.415 & 0.491 & 0.720 & 0.882 & {\bf 0.911} & 0.893 & 0.717 \\ 
  CVB-HS & 0.719 & 0.647 & 0.592 & 0.646 & 0.733 & 0.844 & {\bf 0.957} & {\bf 0.962} & {\bf 0.956} & 0.852 \\ 
  VB-HS & 0.159 & 0.226 & 0.232 & 0.223 & 0.158 & 0.247 & 0.527 & 0.613 & 0.536 & 0.230 \\ 
  MCMC-Lap & 0.604 & 0.528 & 0.521 & 0.529 & 0.606 & 0.742 & 0.890 & {\bf 0.919} & {\bf 0.906} & 0.737 \\ 
  CVB-Lap & 1.390 & 0.792 & 0.598 & 0.797 & 1.448 & {\bf 0.990} & {\bf 0.976} & {\bf 0.952} & {\bf 0.984} & {\bf 0.991} \\ 
  VB-Lap & 0.252 & 0.312 & 0.327 & 0.317 & 0.256 & 0.383 & 0.671 & 0.741 & 0.687 & 0.380 \\ 
\bottomrule
\end{tabular}}
\end{center}
\label{MVS_table_MSEMADMCIWCP}
\end{table}

\begin{table}[thbp]
\caption{Average values of MSE, MAD, MCIW and CP based on $100$ replications for multiple observations (GP) with $k=2$. The minimum values and second smallest values of MSE and MAD are represented in bold and italics respectively. The CP values above 90\% are represented in bold. }
\begin{center}
\resizebox{1.0\textwidth}{!}{ 
\begin{tabular}{c|ccccc|ccccc}
\toprule
& \multicolumn{5}{c|}{MSE} &  \multicolumn{5}{c}{MAD} \\
\hline
(I) Gauss & 0.05 & 0.25 & 0.5 & 0.75 & 0.95 & 0.05 & 0.25 & 0.5 & 0.75 & 0.95 \\ 
  \hline
MCMC-HS & {\bf 0.0188} & {\bf 0.0079} & {\bf 0.0065} & {\bf 0.0076} & {\bf 0.0204} & {\bf 0.1035} & {\bf 0.0693} & {\bf 0.0635} & {\bf 0.0676} & {\bf 0.1080} \\ 
  VB-HS & 0.0267 & 0.0109 & 0.0085 & 0.0102 & 0.0264 & 0.1189 & 0.0791 & 0.0718 & 0.0765 & 0.1231 \\ 
  MCMC-Lap & {\sl 0.0206} & {\sl 0.0083} & {\sl 0.0069} & {\sl 0.0080} & {\sl 0.0224} & {\sl 0.1094} & {\sl 0.0703} & {\sl 0.0646} & {\sl 0.0688} & {\sl 0.1136} \\ 
  VB-Lap & 0.0222 & 0.0089 & 0.0074 & 0.0086 & 0.0234 & 0.1122 & 0.0729 & 0.0669 & 0.0713 & 0.1168 \\ 
  Spline & 0.0394 & 0.0140 & 0.0110 & 0.0135 & 0.0419 & 0.1547 & 0.0915 & 0.0817 & 0.0900 & 0.1605 \\ 
\midrule
\midrule
(II) Beta & 0.05 & 0.25 & 0.5 & 0.75 &0.95 & 0.05 & 0.25 & 0.5 & 0.75 &0.95  \\
\hline
MCMC-HS & 0.0011 & {\bf 0.0014} & {\bf 0.0023} & {\bf 0.0035} & {\bf 0.0084} & 0.0219 & {\sl 0.0234} & {\sl 0.0329} & {\sl 0.0434} & {\bf 0.0702} \\ 
  VB-HS & {\sl 0.0010} & 0.0021 & 0.0032 & 0.0050 & 0.0095 & {\sl 0.0161} & 0.0262 & 0.0375 & 0.0510 & 0.0755 \\ 
  MCMC-Lap & {\bf 0.0008} & {\bf 0.0014} & {\bf 0.0023} & {\bf 0.0035} & 0.0094 & 0.0201 & {\bf 0.0225} & {\bf 0.0321} & {\bf 0.0433} & {\sl 0.0751} \\ 
  VB-Lap & {\bf 0.0008} & {\sl 0.0016} & {\sl 0.0025} & {\sl 0.0039} & {\sl 0.0093} & {\bf 0.0139} & 0.0241 & 0.0339 & 0.0457 & 0.0752 \\ 
  Spline & {\sl 0.0010} & 0.0018 & 0.0028 & 0.0048 & 0.0139 & 0.0171 & 0.0267 & 0.0370 & 0.0515 & 0.0933 \\ 
 \midrule
\midrule
(III) {\small Mixed normal} & 0.05 & 0.25 & 0.5 & 0.75 &0.95 & 0.05 & 0.25 & 0.5 & 0.75 &0.95  \\
\hline
MCMC-HS & {\bf 0.0476} & {\bf 0.019}2 & {\bf 0.0155} & {\bf 0.0180} & {\bf 0.0427} & {\bf 0.1656} & {\bf 0.1084} & {\bf 0.0975} & {\bf 0.1064} & {\bf 0.1630} \\ 
  VB-HS & 0.0900 & 0.0228 & 0.0181 & 0.0223 & 0.0538 & 0.2018 & 0.1181 & 0.1060 & 0.1175 & 0.1810 \\ 
  MCMC-Lap & {\sl 0.0524} & {\sl 0.0197} & {\sl 0.0158} & {\sl 0.0187} & 0.0480 & {\sl 0.1747} & {\sl 0.1102} & {\sl 0.0989} & {\sl 0.1086} & 0.1731 \\ 
  VB-Lap & 0.0672 & 0.0206 & 0.0165 & 0.0199 & {\sl 0.0478} & 0.1838 & 0.1130 & 0.1013 & 0.1120 & {\sl 0.1723} \\ 
  Spline & 0.0925 & 0.0332 & 0.0265 & 0.0331 & 0.0917 & 0.2421 & 0.1453 & 0.1287 & 0.1446 & 0.2422 \\ 
\midrule
\multicolumn{11}{c}{}\\
\midrule
& \multicolumn{5}{c|}{MCIW} &  \multicolumn{5}{c}{CP} \\
\hline
(I) Gauss & 0.05 & 0.25 & 0.5 & 0.75 & 0.95 & 0.05 & 0.25 & 0.5 & 0.75 & 0.95 \\ 
  \hline
MCMC-HS & 0.291 & 0.281 & 0.279 & 0.281 & 0.295 & 0.727 & 0.895 & {\bf 0.914} & 0.899 & 0.715 \\ 
  CVB-HS & 1.077 & 0.551 & 0.374 & 0.566 & 1.086 & {\bf 0.995} & {\bf 0.989} & {\bf 0.961} & {\bf 0.995} & {\bf 0.996} \\ 
  VB-HS & 0.103 & 0.153 & 0.160 & 0.153 & 0.106 & 0.280 & 0.569 & 0.629 & 0.590 & 0.286 \\ 
  MCMC-Lap & 0.305 & 0.294 & 0.294 & 0.294 & 0.308 & 0.732 & {\bf 0.907} & {\bf 0.930} & {\bf 0.911} & 0.723 \\ 
  CVB-Lap & 1.065 & 0.562 & 0.430 & 0.585 & 1.102 & {\bf 0.995} & {\bf 0.994} & {\bf 0.987} & {\bf 0.996} & {\bf 0.996} \\ 
  VB-Lap & 0.131 & 0.181 & 0.192 & 0.181 & 0.133 & 0.364 & 0.687 & 0.752 & 0.707 & 0.369 \\ 
\midrule
\midrule
(II) Beta & 0.05 & 0.25 & 0.5 & 0.75 &0.95 & 0.05 & 0.25 & 0.5 & 0.75 &0.95  \\
\hline
MCMC-HS & 0.074 & 0.103 & 0.133 & 0.157 & 0.170 & {\bf 0.944} & {\bf 0.926} & {\bf 0.907} & 0.852 & 0.665 \\ 
  CVB-HS & 0.524 & 0.252 & 0.185 & 0.375 & 0.743 & {\bf 1.000} & {\bf 0.990} & {\bf 0.943} & {\bf 0.990} & {\bf 0.997} \\ 
  VB-HS & 0.029 & 0.063 & 0.079 & 0.085 & 0.058 & 0.664 & 0.745 & 0.652 & 0.527 & 0.249 \\ 
  MCMC-Lap & 0.072 & 0.104 & 0.137 & 0.162 & 0.179 & {\bf 0.950} & {\bf 0.940} & {\bf 0.921} & 0.871 & 0.661 \\ 
  CVB-Lap & 0.529 & 0.258 & 0.192 & 0.379 & 0.760 & {\bf 1.000} & {\bf 0.992} & {\bf 0.959} & {\bf 0.994} & {\bf 0.997} \\ 
  VB-Lap & 0.036 & 0.072 & 0.092 & 0.099 & 0.075 & 0.806 & 0.824 & 0.762 & 0.638 & 0.311 \\ 
 \midrule
\midrule
(III) {\small Mixed normal} & 0.05 & 0.25 & 0.5 & 0.75 &0.95 & 0.05 & 0.25 & 0.5 & 0.75 &0.95  \\
\hline
MCMC-HS & 0.449 & 0.422 & 0.416 & 0.420 & 0.428 & 0.707 & 0.867 & {\bf 0.907} & 0.883 & 0.696 \\ 
  CVB-HS & 1.327 & 0.771 & 0.547 & 0.769 & 1.290 & {\bf 0.989} & {\bf 0.988} & {\bf 0.958} & {\bf 0.987} & {\bf 0.995} \\ 
  VB-HS & 0.161 & 0.220 & 0.233 & 0.221 & 0.158 & 0.275 & 0.541 & 0.614 & 0.540 & 0.266 \\ 
  MCMC-Lap & 0.478 & 0.441 & 0.436 & 0.439 & 0.461 & 0.712 & 0.881 & {\bf 0.921} & 0.896 & 0.703 \\ 
  CVB-Lap & 1.346 & 0.810 & 0.656 & 0.818 & 1.308 & {\bf 0.991} & {\bf 0.995} & {\bf 0.988} & {\bf 0.995} & {\bf 0.996} \\ 
  VB-Lap & 0.198 & 0.264 & 0.278 & 0.264 & 0.198 & 0.348 & 0.643 & 0.721 & 0.644 & 0.349 \\ 
\bottomrule
\end{tabular}}
\end{center}
\label{MGP_table_MSEMADMCIWCP}
\end{table}

\subsection{More locations}

In Section 4 of the main manuscript, we set the number of locations to $n=100$. We considered the same simulation scenarios as those of Section 4, but we set $n=200$ instead of $n=100$. We note that if we change the number of locations, the (GP) function presented in Subsection \ref{supp:sec:3.1} changes due to the random sampling. To avoid the problem, we considered a linear completion between the original data points. Furthermore, we only compared the following four methods: MCMC-HS, CVB-HS, VB-HS, and ADMM. 

The results are shown in Table~\ref{PC200_table_MSEMADMCIWCP}, \ref{VS200_table_MSEMADMCIWCP} and \ref{GP200_table_MSEMADMCIWCP}. The MSE and MAD for all methods were improved over the results of a smaller sample size in most cases. The point estimation of the MCMC-HS method performed well in many cases as well as the results of $n=100$ observations in terms of point estimation. The coverage probabilities of the MCMC-HS method was also similar to those of $n=100$ observations, which were under 0.90 for extreme quantiles such as 0.05 and 0.95. However, the CVB-HS method provided values over 0.90 except for 0.5 quantile level under (PC) and Beta.

\begin{table}[thbp]
\caption{Average values of MSE, MAD, MCIW and CP based on $100$ replications for 200 locations (PC) with $k=0$. The minimum values and second smallest values of MSE and MAD are represented in bold and italics respectively. The CP values above 90\% are represented in bold. }
\begin{center}
\resizebox{1.0\textwidth}{!}{ 
\begin{tabular}{c|ccccc|ccccc}
\toprule
& \multicolumn{5}{c|}{MSE} &  \multicolumn{5}{c}{MAD } \\
\hline
(I) Gauss & 0.05 & 0.25 & 0.5 & 0.75 & 0.95 & 0.05 & 0.25 & 0.5 & 0.75 & 0.95 \\ 
  \hline
MCMC-HS & 0.0397 & {\bf 0.0060} & {\bf 0.0040} & {\bf 0.0062} & {\bf 0.0404} & 0.1528 & {\bf 0.0547} & {\bf 0.0459} & {\bf 0.0565} & {\bf 0.1542} \\ 
  VB-HS & {\sl 0.0358} & {\sl 0.0176} & {\sl 0.0122} & {\sl 0.0172} & {\sl 0.0375} & {\sl 0.1479} & 0.0934 & 0.0765 & 0.0921 & 0.1511 \\ 
  ADMM & {\bf 0.0320} & 0.0238 & 0.0143 & 0.0202 & 0.0366 & {\bf 0.1309} & 0.1187 & 0.0912 & 0.1111 & 0.1335 \\ 
\midrule
\midrule
(II) Beta & 0.05 & 0.25 & 0.5 & 0.75 &0.95 & 0.05 & 0.25 & 0.5 & 0.75 &0.95  \\
\hline
MCMC-HS & 0.0052 & {\bf 0.0016} & {\bf 0.0024} & {\bf 0.0042} & 0.0209 & 0.0214 & {\bf 0.0200} & {\bf 0.0310} & {\bf 0.0447} & 0.1112 \\ 
  VB-HS & {\bf 0.0005} & 0.0026 & {\sl 0.0043} & {\sl 0.0077} & {\bf 0.0144} & {\bf 0.0100} & {\sl 0.0265} & {\sl 0.0404} & {\sl 0.0601} & {\bf 0.0944} \\ 
  ADMM & {\sl 0.0009} & {\sl 0.0021} & 0.0045 & 0.0130 & {\sl 0.0185} & {\sl 0.0130} & 0.0344 & 0.0514 & 0.0927 & {\sl 0.097}5 \\  
 \midrule
\midrule
(III) {\small Mixed normal} & 0.05 & 0.25 & 0.5 & 0.75 &0.95 & 0.05 & 0.25 & 0.5 & 0.75 &0.95  \\
\hline
MCMC-HS & {\bf 0.0777} & {\bf 0.0201} & {\bf 0.0144} & {\bf 0.0181} & {\bf 0.0811} & {\sl 0.2202} & {\bf 0.0958} & {\bf 0.0824} & {\bf 0.0942} & {\sl 0.2234} \\ 
  VB-HS & {\sl 0.0866} & {\sl 0.0427} & {\sl 0.0325} & {\sl 0.0440} & {\sl 0.0898} & 0.2296 & {\sl 0.1510} & {\sl 0.1274} & {\sl 0.1530} & 0.2309 \\ 
  ADMM & 0.0891 & 0.0638 & 0.0471 & 0.0590 & 0.0940 & {\bf 0.2184} & 0.1879 & 0.1596 & 0.1830 & {\bf 0.2162} \\ 
\midrule
\multicolumn{11}{c}{}\\
\midrule
& \multicolumn{5}{c|}{MCIW} &  \multicolumn{5}{c}{CP} \\
\hline
(I) Gauss & 0.05 & 0.25 & 0.5 & 0.75 & 0.95 & 0.05 & 0.25 & 0.5 & 0.75 & 0.95 \\ 
  \hline
MCMC-HS & 0.474 & 0.286 & 0.251 & 0.280 & 0.477 & 0.746 & {\bf 0.942} & {\bf 0.957} & {\bf 0.927} & 0.746 \\ 
  CVB-HS & 1.003 & 0.507 & 0.374 & 0.497 & 0.991 & {\bf 0.973} & {\bf 0.967} & {\bf 0.943} & {\bf 0.962} & {\bf 0.968} \\ 
  VB-HS & 0.115 & 0.187 & 0.194 & 0.185 & 0.117 & 0.237 & 0.619 & 0.724 & 0.625 & 0.235 \\ 
\midrule
\midrule
(II) Beta & 0.05 & 0.25 & 0.5 & 0.75 &0.95 & 0.05 & 0.25 & 0.5 & 0.75 &0.95  \\
\hline
MCMC-HS & 0.066 & 0.094 & 0.136 & 0.185 & 0.270 & {\bf 0.976} & {\bf 0.948} & {\bf 0.920} & 0.895 & 0.625 \\ 
  CVB-HS & 0.359 & 0.229 & 0.161 & 0.327 & 0.662 & {\bf 0.997} & {\bf 0.975} & 0.892 & {\bf 0.947} & {\bf 0.977} \\ 
  VB-HS & 0.032 & 0.071 & 0.090 & 0.098 & 0.062 & 0.871 & 0.822 & 0.720 & 0.566 & 0.207 \\ 
 \midrule
\midrule
(III) {\small Mixed normal} & 0.05 & 0.25 & 0.5 & 0.75 &0.95 & 0.05 & 0.25 & 0.5 & 0.75 &0.95  \\
\hline
MCMC-HS & 0.758 & 0.463 & 0.425 & 0.455 & 0.758 & 0.778 & {\bf 0.924} & {\bf 0.949} & {\bf 0.927} & 0.778 \\ 
  CVB-HS & 1.435 & 0.766 & 0.629 & 0.756 & 1.449 & {\bf 0.979} & {\bf 0.968} & {\bf 0.949} & {\bf 0.962} & {\bf 0.975} \\ 
  VB-HS & 0.177 & 0.281 & 0.293 & 0.282 & 0.177 & 0.232 & 0.564 & 0.672 & 0.560 & 0.237 \\ 
\bottomrule
\end{tabular}}
\end{center}
\label{PC200_table_MSEMADMCIWCP}
\end{table}

\begin{table}[thbp]
\caption{Average values of MSE, MAD, MCIW and CP based on $100$ replications for 200 locations (VS) with $k=1$. The minimum values and second smallest values of MSE and MAD are represented in bold and italics respectively. The CP values above 90\% are represented in bold. }
\begin{center}
\resizebox{1.0\textwidth}{!}{ 
\begin{tabular}{c|ccccc|ccccc}
\toprule
& \multicolumn{5}{c|}{MSE} &  \multicolumn{5}{c}{MAD} \\
\hline
(I) Gauss & 0.05 & 0.25 & 0.5 & 0.75 & 0.95 & 0.05 & 0.25 & 0.5 & 0.75 & 0.95 \\ 
  \hline
MCMC-HS & {\bf 0.0279} & {\bf 0.0101} & {\bf 0.0090} & {\bf 0.0104} & {\bf 0.0279} & {\bf 0.1264} & {\bf 0.0759} & {\bf 0.0701} & {\bf 0.0751} & {\bf 0.1240} \\ 
  VB-HS & {\sl 0.0365} & {\sl 0.0147} & {\sl 0.0124} & {\sl 0.0162} & {\sl 0.0362} & {\sl 0.1449} & {\sl 0.0889} & {\sl 0.0810} & {\sl 0.0921} & {\sl 0.1439} \\ 
  ADMM & 0.0719 & 0.0151 & 0.0138 & 0.0166 & 0.0405 & 0.1659 & 0.0927 & 0.0880 & 0.0968 & 0.1488 \\ 
\midrule
\midrule
(II) Beta & 0.05 & 0.25 & 0.5 & 0.75 &0.95 & 0.05 & 0.25 & 0.5 & 0.75 &0.95  \\
\hline
MCMC-HS & {\sl 0.0023} & {\bf 0.0019} & {\bf 0.0027} & {\bf 0.0040} & {\bf 0.0115} & {\sl 0.0300} & {\bf 0.0277} & {\bf 0.0365} & {\bf 0.0460} & {\bf 0.0812} \\ 
  VB-HS & {\bf 0.0022} & 0.0025 & 0.0038 & 0.0060 & {\sl 0.0134} & {\bf 0.0244} & {\sl 0.0300} & {\sl 0.0419} & 0.0559 & {\sl 0.0884} \\ 
  ADMM & 0.0946 & {\sl 0.0023} & {\sl 0.0037} & {\sl 0.0055} & 0.0223 & 0.1058 & 0.0311 & 0.0427 & {\sl 0.0552} & 0.1010 \\ 
 \midrule
\midrule
(III) {\small Mixed normal} & 0.05 & 0.25 & 0.5 & 0.75 &0.95 & 0.05 & 0.25 & 0.5 & 0.75 &0.95  \\
\hline
MCMC-HS & {\bf 0.0622} & {\bf 0.0269} & {\bf 0.0204} & {\bf 0.0237} & {\bf 0.0576} & {\bf 0.1858} & {\bf 0.1207} & {\bf 0.1080} & {\bf 0.1165} & {\bf 0.1854} \\ 
  VB-HS & {\sl 0.0750} & {\sl 0.0310} & {\sl 0.0241} & {\sl 0.0316} & 0.0795 & {\sl 0.2116} & {\sl 0.1337} & {\sl 0.1175} & {\sl 0.1337} & 0.2195 \\ 
  ADMM & 0.0950 & 0.0336 & 0.0289 & 0.0356 & {\sl 0.0732} & 0.2168 & 0.1416 & 0.1322 & 0.1456 & {\sl 0.2064} \\ 
\midrule
\multicolumn{11}{c}{}\\
\midrule
& \multicolumn{5}{c|}{MCIW} &  \multicolumn{5}{c}{CP} \\
\hline
(I) Gauss & 0.05 & 0.25 & 0.5 & 0.75 & 0.95 & 0.05 & 0.25 & 0.5 & 0.75 & 0.95 \\ 
  \hline
MCMC-HS & 0.362 & 0.321 & 0.315 & 0.322 & 0.354 & 0.725 & {\bf 0.903} & {\bf 0.923} & {\bf 0.904} & 0.732 \\ 
  CVB-HS & 1.060 & 0.522 & 0.418 & 0.506 & 1.045 & {\bf 0.988} & {\bf 0.973} & {\bf 0.943} & {\bf 0.955} & {\bf 0.982} \\ 
  VB-HS & 0.120 & 0.187 & 0.192 & 0.186 & 0.121 & 0.264 & 0.625 & 0.692 & 0.618 & 0.280 \\ 
\midrule
\midrule
(II) Beta & 0.05 & 0.25 & 0.5 & 0.75 &0.95 & 0.05 & 0.25 & 0.5 & 0.75 &0.95  \\
\hline
MCMC-HS & 0.098 & 0.125 & 0.151 & 0.175 & 0.212 & 0.927 & {\bf 0.935} & {\bf 0.905} & 0.874 & 0.683 \\ 
  CVB-HS & 0.460 & 0.251 & 0.187 & 0.332 & 0.677 & {\bf 0.998} & {\bf 0.982} & {\bf 0.920} & {\bf 0.962} & {\bf 0.986} \\ 
  VB-HS & 0.038 & 0.079 & 0.097 & 0.103 & 0.066 & 0.692 & 0.784 & 0.698 & 0.583 & 0.246 \\ 
 \midrule
\midrule
(III) {\small Mixed normal} & 0.05 & 0.25 & 0.5 & 0.75 &0.95 & 0.05 & 0.25 & 0.5 & 0.75 &0.95  \\
\hline
MCMC-HS & 0.515 & 0.490 & 0.479 & 0.477 & 0.540 & 0.711 & 0.890 & {\bf 0.918} & 0.891 & 0.745 \\ 
  CVB-HS & 1.315 & 0.721 & 0.637 & 0.690 & 1.383 & {\bf 0.984} & {\bf 0.963} & {\bf 0.962} & {\bf 0.953} & {\bf 0.985} \\ 
  VB-HS & 0.182 & 0.268 & 0.281 & 0.267 & 0.182 & 0.272 & 0.589 & 0.676 & 0.593 & 0.263 \\ 
\bottomrule
\end{tabular}}
\end{center}
\label{VS200_table_MSEMADMCIWCP}
\end{table}

\begin{table}[thbp]
\caption{Average values of MSE, MAD, MCIW and CP based on $100$ replications for 200 locations (GP) with $k=2$. The minimum values and second smallest values of MSE and MAD are represented in bold and italics respectively. The CP values above 90\% are represented in bold. }
\begin{center}
\resizebox{1.0\textwidth}{!}{ 
\begin{tabular}{c|ccccc|ccccc}
\toprule
& \multicolumn{5}{c|}{MSE} &  \multicolumn{5}{c}{MAD} \\
\hline
(I) Gauss & 0.05 & 0.25 & 0.5 & 0.75 & 0.95 & 0.05 & 0.25 & 0.5 & 0.75 & 0.95 \\ 
  \hline
MCMC-HS & {\bf 0.0329} & {\bf 0.0111} & {\bf 0.0094} & {\bf 0.0115} & {\bf 0.0277} & {\bf 0.1309} & {\bf 0.0812} & {\bf 0.0747} & {\bf 0.0815} & {\bf 0.1239} \\ 
  VB-HS & 0.1083 & 0.0131 & {\sl 0.0107} & {\sl 0.0142} & {\sl 0.0349} & 0.1854 & {\sl 0.0877} & {\sl 0.0788} & {\sl 0.0885} & {\sl 0.1408} \\ 
  ADMM & {\sl 0.0409} & {\sl 0.0199} & 0.0180 & 0.0218 & 0.0397 & {\sl 0.1531} & 0.1077 & 0.1017 & 0.1108 & 0.1516 \\
\midrule
\midrule
(II) Beta & 0.05 & 0.25 & 0.5 & 0.75 &0.95 & 0.05 & 0.25 & 0.5 & 0.75 &0.95  \\
\hline
MCMC-HS & {\sl 0.0016} & {\bf 0.0017} & {\bf 0.0028} & {\bf 0.0044} & {\bf 0.0100} & 0.0271 & {\bf 0.0264} & {\bf 0.0373} & {\bf 0.0491} & {\bf 0.0777} \\ 
  VB-HS & {\bf 0.0015} & {\sl 0.0023} & {\sl 0.0034} & {\sl 0.0054} & {\sl 0.0120} & {\bf 0.0206} & {\sl 0.0284} & {\sl 0.0406} & {\sl 0.0537} & {\sl 0.0854} \\ 
  ADMM & {\bf 0.0015} & 0.0028 & 0.0047 & 0.0068 & 0.0133 & 0.0209 & 0.0330 & 0.0478 & 0.0613 & 0.0904 \\ 
 \midrule
\midrule
(III) {\small Mixed normal} & 0.05 & 0.25 & 0.5 & 0.75 &0.95 & 0.05 & 0.25 & 0.5 & 0.75 &0.95  \\
\hline
MCMC-HS & {\sl 0.1091} & {\bf 0.0337} & {\bf 0.0231} & {\bf 0.0265} & {\bf 0.0615} & {\bf 0.2281} & {\bf 0.1367} & {\bf 0.1177} & {\bf 0.1272} & {\bf 0.1936} \\ 
  VB-HS & 0.1447 & {\sl 0.0356} & {\bf 0.0231} & {\sl 0.0276} & {\sl 0.0770} & 0.2562 & {\sl 0.1375} & {\sl 0.1180} & {\sl 0.1299} & {\sl 0.2163} \\ 
  ADMM & {\bf 0.0911} & 0.0471 & {\sl 0.0423} & 0.0480 & 0.0939 & {\sl 0.2400} & 0.1719 & 0.1618 & 0.1715 & 0.2394 \\ 
\midrule
\multicolumn{11}{c}{}\\
\midrule
& \multicolumn{5}{c|}{MCIW} &  \multicolumn{5}{c}{CP} \\
\hline
(I) Gauss & 0.05 & 0.25 & 0.5 & 0.75 & 0.95 & 0.05 & 0.25 & 0.5 & 0.75 & 0.95 \\ 
  \hline
MCMC-HS & 0.368 & 0.358 & 0.353 & 0.355 & 0.338 & 0.723 & {\bf 0.914} & {\bf 0.937} & {\bf 0.917} & 0.734 \\ 
  CVB-HS & 1.105 & 0.528 & 0.440 & 0.522 & 1.065 & {\bf 0.964} & {\bf 0.978} & {\bf 0.966} & {\bf 0.972} & {\bf 0.987} \\ 
  VB-HS & 0.139 & 0.195 & 0.206 & 0.196 & 0.130 & 0.280 & 0.632 & 0.709 & 0.642 & 0.300 \\ 
\midrule
\midrule
(II) Beta & 0.05 & 0.25 & 0.5 & 0.75 &0.95 & 0.05 & 0.25 & 0.5 & 0.75 &0.95  \\
\hline
MCMC-HS & 0.096 & 0.126 & 0.160 & 0.188 & 0.190 & {\bf 0.947} & {\bf 0.948} & {\bf 0.913} & 0.875 & 0.670 \\ 
  CVB-HS & 0.462 & 0.252 & 0.204 & 0.333 & 0.693 & {\bf 0.999} & {\bf 0.986} & {\bf 0.945} & {\bf 0.968} & {\bf 0.990} \\ 
  VB-HS & 0.037 & 0.081 & 0.100 & 0.105 & 0.072 & 0.635 & 0.803 & 0.705 & 0.596 & 0.267 \\ 
 \midrule
\midrule
(III) {\small Mixed normal} & 0.05 & 0.25 & 0.5 & 0.75 &0.95 & 0.05 & 0.25 & 0.5 & 0.75 &0.95  \\
\hline
MCMC-HS & 0.5628 & 0.5995 & 0.5890 & 0.5730 & 0.5322 & 0.6995 & {\bf 0.9126} & {\bf 0.9469} & {\bf 0.9162} & 0.7287 \\ 
  CVB-HS & 1.393 & 0.753 & 0.654 & 0.720 & 1.404 & {\bf 0.962} & {\bf 0.967} & {\bf 0.965} & {\bf 0.963} & {\bf 0.987} \\ 
  VB-HS & 0.198 & 0.283 & 0.300 & 0.281 & 0.194 & 0.260 & 0.595 & 0.690 & 0.615 & 0.278 \\ 
\bottomrule
\end{tabular}}
\end{center}
\label{GP200_table_MSEMADMCIWCP}
\end{table}

\section{Additional information on real data example}

\begin{figure}[thbp]
\begin{center}
\includegraphics[width=\linewidth]{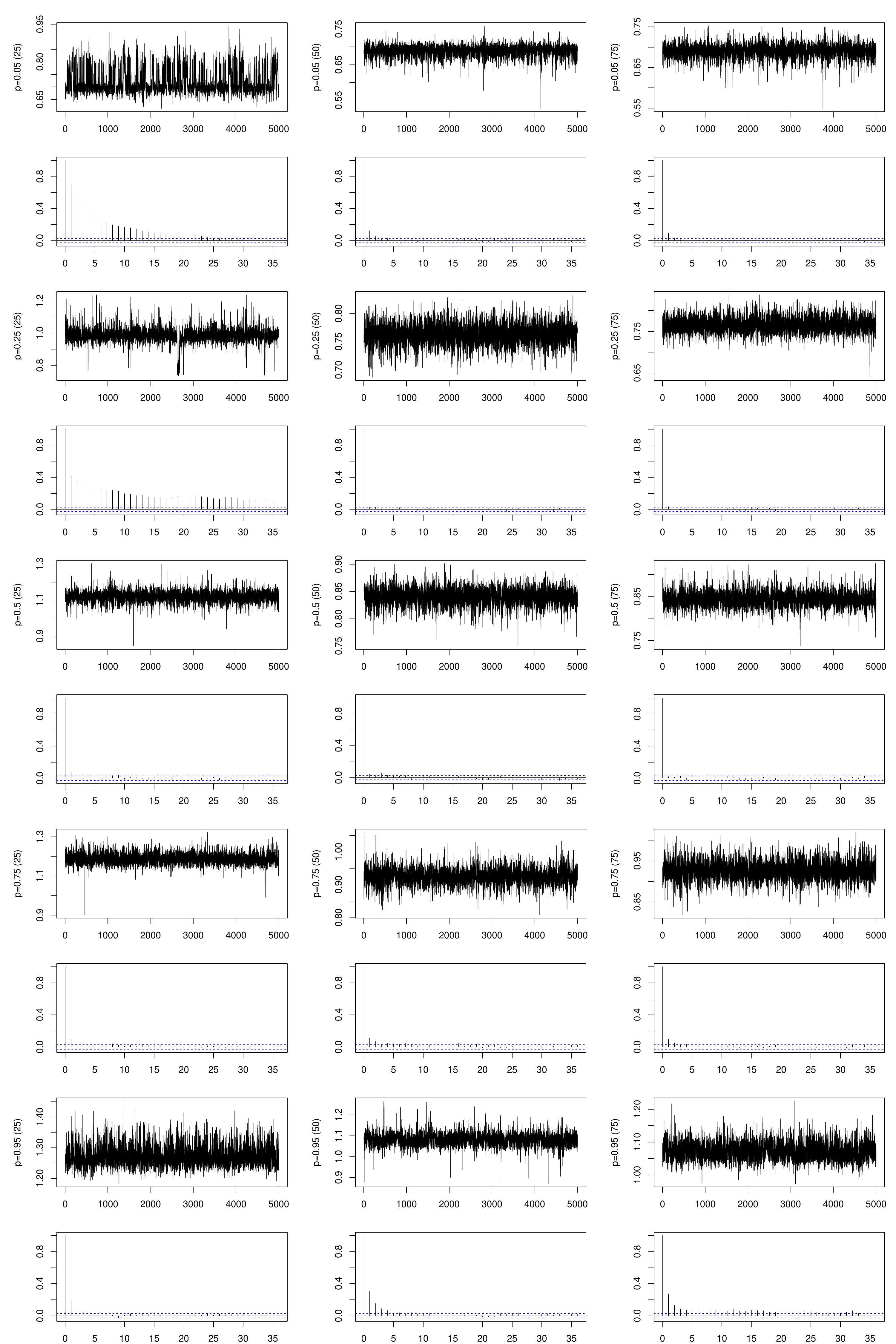}
\caption{Trace plots and autocorrelations of posterior samples of $\theta_{25}, \theta_{50}, \theta_{75}$ (from left to right) for quantile levels $0.05, 0.25, 0.50, 0.75, 0.95$ (from top to bottom) in real data analysis of Nile data.}
\label{Nile_sample_acf}
\end{center}
\end{figure}

\begin{figure}[thbp]
\begin{center}
\includegraphics[width=\linewidth]{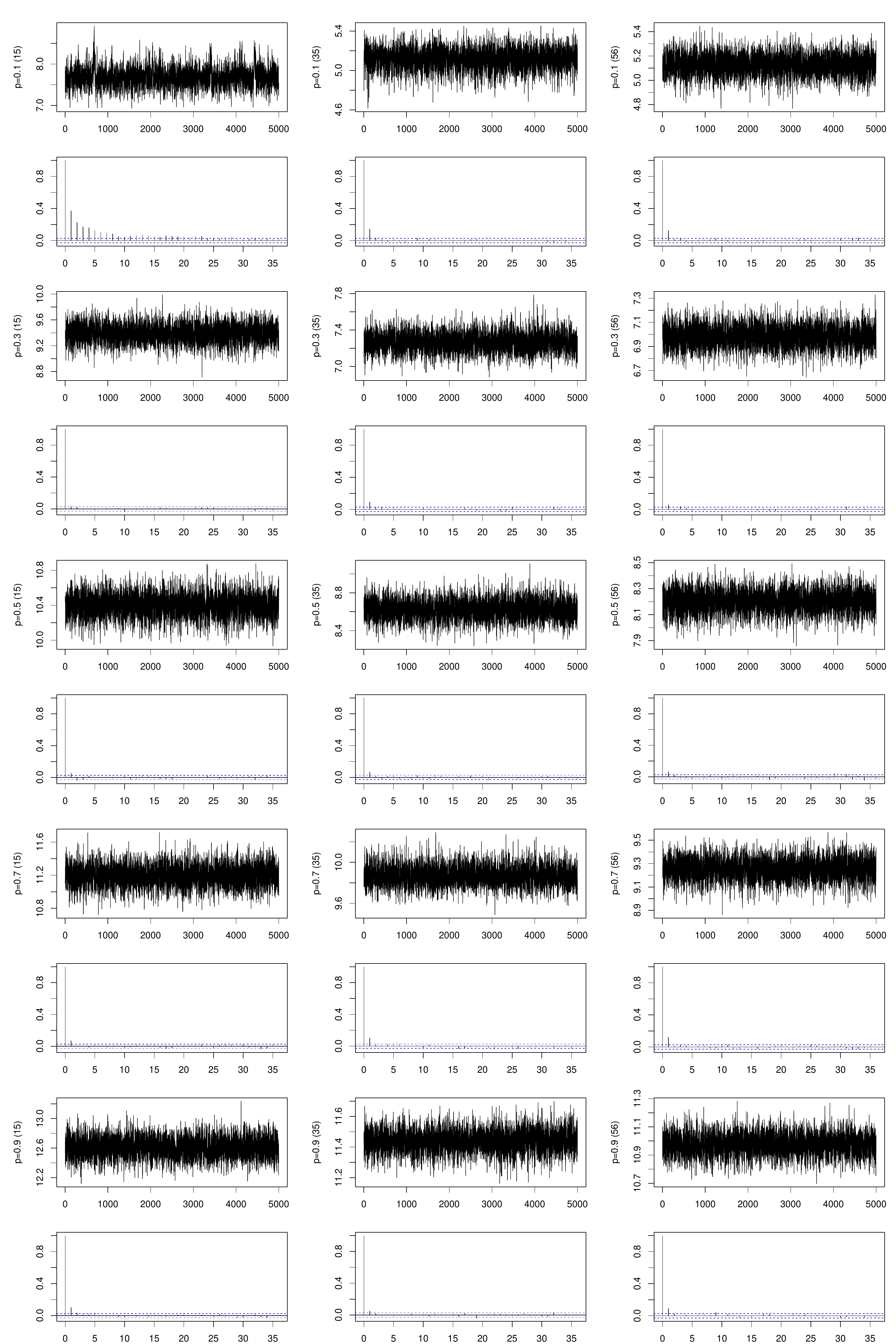}
\caption{Trace plots and autocorrelations of posterior samples of $\theta_{15}, \theta_{35}, \theta_{56}$ (from left to right) for quantile levels $0.10, 0.30, 0.50, 0.70, 0.90$ (from top to bottom) in real data analysis of Munich rent data.}
\label{munich_sample_acf}
\end{center}
\end{figure}

We confirm that the proposed algorithm works well by making a simple diagnosis of sampling efficiency. Here, we especially consider the sampling efficiency in the two real data analyses. We generate 60,000 posterior samples under MCMC in section 2.3 after discarding the first 10,000 posterior samples as burn-in, and then only every 10th scan was saved. For Nile data in Section 5.1 and Munich rent data in Section 5.2, sample paths and autocorrelations of the posterior samples in three selected points are provided in Figures \ref{Nile_sample_acf} and \ref{munich_sample_acf}, respectively. The mixing and autocorrelation of the MCMC algorithm seem satisfactory.

\end{document}